\documentclass[12pt,preprint]{aastex}

\begin{document}

\title{Simultaneous Multiwavelength and Optical Microvariability Observations of CTA 102 (PKS J2232+1143)}

\shorttitle{CTA 102; Observations}
\shortauthors{Osterman Meyer et al.}

\author{Angela Osterman Meyer\altaffilmark{a}, H. Richard Miller\altaffilmark{b}, Kevin Marshall\altaffilmark{c}, Wesley T. Ryle\altaffilmark{d}, Hugh Aller\altaffilmark{e}, Margo Aller\altaffilmark{e}, and Tom Balonek\altaffilmark{f} }

\altaffiltext{a}{Evelyn L. Egan Observatory at Florida Gulf Coast University, College of Arts and Sciences, 10501 FGCU Blvd, S., Fort Myers, FL, 33965, USA; ameyer@fgcu.edu}

\altaffiltext{b}{Georgia State University, Dept. of Physics and Astronomy, 1 Park Place ste. 730, Atlanta, GA, 30303, USA}

\altaffiltext{c}{Bucknell University, Dept. of Physics and Astronomy, 701 Moore Avenue, Lewisburg, PA, 17837, USA}

\altaffiltext{d}{Thomas More College, Dept. of Math and Physics, 333 Thomas More Parkway, Crestview Hills, KY, 41017, USA}

\altaffiltext{e}{University of Michigan, Dept. of Astronomy, 500 Church St. 830 Dennison, Ann Arbor, MI, 48109-1042, USA}

\altaffiltext{f}{Foggy Bottom Observatory at Colgate University, Hamilton, NY, USA}

\begin{abstract}
We present analysis of both the short-term optical and long-term multiwavelength variability of CTA 102.
In 2004, this object was observed in an intense optical flaring state.
Extensive R-band microvariability observations were carried out during this high state.
In 2005, we obtained several weeks of contemporaneous radio, optical, and X-ray observations of CTA 102.
These observations recorded distinct flaring activity in all three wavebands.
Subsequent analysis revealed that this object may appear redder when in a brighter optical state, and that the X-ray, optical and radio activity do not appear to be correlated.
The shape of the observed spectral energy distributions suggests that both synchrotron-related and external inverse Compton processes may contribute to the X-ray emission.
Our results are also compared to other results on this object and archival microvariability observations.
It appears that more rapid, dramatic microvariability events occur when CTA 102 is in an elevated optical flux state.
\end{abstract}

\keywords{galaxies: active --- galaxies: individual(\objectname{CTA 102})}

\section{Introduction}

Blazars are a radio-loud subclass of active galactic nuclei (AGNs), typically featuring extreme core-dominated radio morphologies. 
Their optical continua are markedly steep, they exhibit a high degree of optical polarization (up to about 20\%), and their fluxes vary on timescales of an hour to several years at all observed wavelengths.
The most extreme and unique property of blazars is their highly beamed continuum, most likely produced by a jet of relativistic material oriented close to the observer's line of sight \citep[][ and references therein]{urr95}.

Blazars can be subdivided into two categories: BL Lac objects and Flat Spectrum Radio Quasars (FSRQs).
BL Lac objects have particularly weak spectral lines, so their redshifts, and hence luminosities, are difficult to determine.
FSRQs are distinguished by the presence of broad optical emission lines.
As the continuum flux varies, the distinction between the two classes may become blurred.
Spectral lines indicating an FSRQ may be present during a low-continuum state, but such lines may become weak or disappear during a high-continuum state, thus resembling a BL Lac object \citep[][ and references therein]{ulr97}.

Blazars exhibit the most extreme variability observed for any class of AGN.
Optical microvariability has been detected on timescales as short as minutes to hours, with similar variations observed at all wavelengths except radio \citep{
mil96}.
The earliest microvariability results were from BL Lac, one of the most frequently studied blazars \citep{rac70,mil89}.
Observations of close correlations between flares observed in different wavebands strongly indicated that these variations are intrinsic to the blazars themselves \citep{wag95}.
Possible extrinsic explanations, such as interstellar scintillation and gravitational microlensing, do not match the observations.

FSRQs and low-frequency peaked BL Lac objects (LBLs) typically exhibit strong optical microvariability on timescales as short as several minutes.
Among recent observations of BL Lac, \citet{bot04} observed $\Delta$R of $\sim$ 0.35 magnitudes (or, in terms of flux F, $\Delta$F/F of $\sim$ 0.4) over 1.5 hours, \citet{nik05} observed a change in R of 0.10 $\pm$ 0.01 magnitudes within 1 hour, and \citet{ben06} observed $\Delta$R of 0.147 $\pm$ 0.020 magnitudes over 2.5 hours.
Observations of 3C 279 from 1989 to 2002 found several instances of R-band microvariability covering many different overall flux states \citep{bal02a,bal02b}.
\citet{how04} observed R-band microvariability in 3C 345.
They also note that there was no apparent correlation between the occurrence of microvariability and the overall flux state of the object.

The spectral energy distribution (SED) of blazars is distinguished by two peaks: one in the radio/UV regime and the other in the X-ray/$\gamma$-ray regime. 
The spectrum in the radio to UV range is generally agreed to arise from synchrotron emission from relativistic electrons spiraling around the jet's magnetic field lines \citep{ulr97}. 
The X-ray/$\gamma$-ray spectrum is most likely due to inverse Compton (IC) emission in FSRQs and LBLs, while most of the X-rays are the high-energy tail of the synchrotron emission in high-frequency peaked BL Lac objects (HBLs) \citep{ulr97}. 
A major unsolved puzzle of blazars is what supplies the ``seed photons" which are upscattered to produce the IC emission. 
In order to fully understand blazars, we must understand the physics of the region near the central engine, since this is where jet particles are collimated and accelerated to relativistic speeds. 

CTA 102, also known as PKS J2232+1143, has a redshift of 1.037 \citep{sch65}, and was first identified as a radio source by the Caltech Radio Survey \citep{har60}.
\citet{san65} confirmed that it was a quasar, and observations of a blue stellar optical counterpart were published the same year \citep{wyn65}.
CTA 102's optical polarization, observed at nearly 11\%, put it in the class of High Polarization Quasars (HPQs), objects which exhibit optical polarization of at least 3\% \citep{moo81}.
Its spectral properties identify it as an optically violent variable (OVV) quasar \citep{mar86}.
Variability in CTA 102 was observed in the radio \citep{hun72} and optical bands \citep[e.g.,][]{pic88} as early as the late 1960s \citep{bar78}.
\citet{tor99} observed a simultaneous radio and optical flare in CTA 102 in 1997.

In this work, we present the results of an intensive optical microvariability campaign performed on CTA 102 during a very high flux state.
The results of this campaign are compared with microvariability observed in this object at other times and different flux levels.
We also present the results of the first simultaneous radio, optical, and X-ray campaign performed for this object.
We discuss the correlations, or absence thereof, found between the various wavebands and the variation in optical color present during an optical flare.
We present and discuss the broadband SEDs and compare them with previous results on CTA 102 and similar objects.

\section{Observations}

The data presented in this work were collected over a wide variety of timescales and from several observing facilities.
In this section, the data reduction and the processing procedures will be discussed.

\subsection{X-ray Data Reduction}

All of the X-ray observations were obtained from the \emph{Rossi X-ray Timing Explorer} (\emph{RXTE}).
The X-ray light curve was extracted using the {\sc FTOOLS}~v5.2 software package.
During nearly all of our observations, Proportional Counting Units (PCUs) 1, 3, and 4 were turned off.
Therefore, data were only extracted from PCUs 0 and 2.
Despite the loss of the propane layer on board PCU 0 during 2000 May, the signal-to-noise ratio was much greater when using data from both PCUs 0 and 2.
To further enhance the signal-to-noise ratio, only data from layer 1 of the Proportional Counter Array (PCA) were analyzed.
No data from the High-Energy X-Ray Timing Experiment (HEXTE) cluster or the other PCUs were used.
All of the data analyzed here were taken while the spacecraft was in {\sc STANDARD-2} data mode.

Data were extracted only when the target's Earth elevation angle was $>10\deg$, pointing offset $<0.02\deg$, PCUs 0 and 2 both on, the spacecraft more than 30 minutes after South Atlantic Anomaly (SAA) passage, and electron noise less than 0.1 units.
Since the background response of the PCU is not well defined above 20~keV, only channels $0-44$ ($2-20$~keV) are included in this analysis.

Because the PCA is a non-imaging detector, background issues can be critically important during analysis.
The faint-mode ``L7'' model, developed by the PCA team, was used here.
This model provides adequate background estimation for objects with less than 40 cnts/sec.
Background files were extracted using {\tt pcabackest}~v3.0.

\subsection{Optical Data Reduction}

All of the Lowell Observatory and Small and Moderate Aperture Research Telescope System (SMARTS) observations were taken, reduced, and/or processed by the Program for Extragalactic Astronomy (PEGA) group at Georgia State University.
Details of the Lowell Observatory and SMARTS telescopes are given in Table 1.
Bias/zero and flat calibration frames were taken along with the CTA 102 object frames.
Dark calibration frames were not required for the Lowell Observatory or SMARTS CCD since the chips contain a dark pixel strip.
The object frames also included a number of comparison stars used to calibrate the observed brightness of CTA 102 using differential photometry.
The information on each comparison star is provided in Table 2, with a labeled optical finder's chart displayed in Figure 1.
All data reduction utilized standard NOAO IRAF\footnote{IRAF is distributed by the National Optical Astronomy Observatories, which are operated by the Association of the Universities for Research in Astronomy, Inc., under cooperative agreement with the National Science Foundation.} routines including {\tt ccdproc}, {\tt flatcombine}, and {\tt zerocombine}.
All data processing and 7 arcsec aperture photometry was done using the {\tt ccdphot} routine, written by Marc Buie, in IDL.

For the 2005 CTA 102 campaign, some data were obtained from the Foggy Bottom Observatory at Colgate University in Hamilton, NY.
These observations were collected using the Ferson 16-inch reflecting telescope and Photometrics PM3000 CCD camera and a Johnson R filter.
The images were then calibrated and reduced using IRAF scripts written by Christy Tremonti \citep{bec97}.
Photometry on the images was performed by Tom Balonek and his students using methods comparable to those of the PEGA group.

\subsection{Radio Data Reduction}

The University of Michigan's Radio Astronomy Observatory (UMRAO) data were obtained using a 26m prime focus paraboloid equipped with transistor-based radiometers operating at central frequencies of 4.8, 8.0 and 14.5 GHz and room-temperature wide-band High Electron Mobility Pseudomorphic Transistor (HEMPT) amplifiers (with a width of $\sim$10\% of the observing frequency).
Measurements at all three frequencies utilized rotating, dual-horn polarimeter feed systems which permitted both total flux density and linear polarization to be measured. 
An on--off observing technique was used at 4.8 GHz, and an on--on technique at the other two frequencies.
A typical observation consisted of 8 to 16 individual measurements over a 25 to 45 minute period (depending on frequency).
A source selected from a grid of calibrators was observed every 1 to 2 hours. 
The flux scale was set by observations of Cassiopeia A. 
Details of the calibration and analysis technique are given in \citet{all85}.

\section{Microvariability Studies}

During a 2004 September observing run on the Lowell Observatory Perkins telescope, a routine observation of CTA 102 was made for long-term light-curve purposes.
This object is usually relatively faint (see Figure 2); normally  much fainter than optical check stars 1 and 2.
We found CTA 102 to be in a bright state, appearing at least as bright as these two check stars and the brightest observed state displayed in Figure 2.
As a result of this detection, we made this object the main focus of that observing run, and of the subsequent Lowell Perkins time in October.
This campaign captured some extremely rapid optical flares and very dramatic changes in brightness, as displayed in Figures 3 and 4.
The observations from September 21 and October 4 display some of the most well-defined, rapid flaring events ever observed.
Early on the night of the 21st, CTA 102 brightens by about 0.06 magnitudes in roughly 15 minutes.
During the latter part of the observations on October 4, one observes a 0.07 magnitude increase over about 15 minutes followed by a symmetrical decrease.
About half an hour later, an increase of 0.04 magnitudes over around 15 minutes is immediately followed by a decrease of 0.09 magnitudes over the same amount of time.

In this work, we also present microvariability observations of CTA 102 from 2002, when the object was in an elevated flux state, though not as bright as during the 2004 campaign.
Significant flaring events observed in 2002 are displayed in Figure 5. 
There are also several nights where no significant microvariability events were detected.
Table 3 summarizes all of the microvariability observations presented in this work.

A quantitative analysis was performed on all of the microvariability observations summarized in Table 3.
We define an event to be an increase or decrease in brightness sustained over at least five data points resulting in at least a 2$\sigma$ change in brightness with respect to the errors determined from the observed differential magnitudes.
Almost every event listed in Table 3 has a significance of at least 2.6$\sigma$.
Table 3 also gives the time-rate-of-change in brightness of each event as the magnitude rate in units of magnitudes per day.
We use the magnitude rates to quantitatively compare how fast the object's brightness is changing.

Figure 6 displays the rate and duration of each microvariability event compared to the observed brightness of CTA 102 on the night the event occurred.
Slower changes in brightness ($\sim$2 magnitudes per day or less) appear to occur over the full range of observed brightnesses in this object.
However, it appears that faster increases and decreases in brightness only occur when the object is in a very elevated flux state.
An interesting trend emerges when examining the duration of flaring events versus object brightness.
In dim and medium brightness states, the microvariations occur in a range of time durations, from one to several hours in length.
However, in the brightest observed states, the significant microvariations observed in CTA 102 all last about a hundredth of a day, or only about 15 minutes.
It appears that in extreme flaring states of CTA 102, the variations become very rapid and occur on very short timescales.
The probability of a microvariability event occurring does not appear to depend on the overall flux state of CTA 102, at least not when the object is in medium to high flux states.
As displayed in Figure 6, the events are more or less evenly distributed across observed R magnitudes, even though the most rapid events only occur in very bright states.

These results can be compared with previous studies of microvariability structure compared to flux state.
\citet{how04} find no correlation between flux state and occurrences of microvariability in blazars, which agrees with our CTA 102 observations during moderate and high flux states.
They do find that more microvariability events occur when the object's long-term flux behavior is undergoing a significant increase or decrease \citep{how04}.
Our observations of many microvariability events during the 2004 campaign just before an overall decrease in brightness appears to agree with this prediction.
However, there is an inherent selection effect present because many nights of data were taken during the 2004 campaign compared to other times.

\section{Multiwavelength Behavior}

Regular optical monitoring of CTA 102 began in 2005 May from Foggy Bottom Observatory and Lowell Observatory.
The magnitude required to trigger \emph{RXTE} Target of Opportunity observations, R of 16.0 or brighter, was reached in late 2005 September.
The \emph{RXTE} campaign extended from 2005 September 24$ - $October 16.
Throughout the \emph{RXTE} campaign, regular optical observations were made through the SMARTS program.
In addition, intensive optical observations, particularly in the R-band, were made from Lowell Observatory in mid-October.
These observations did not contain any well-defined microvariability events like that observed in 2004.

CTA 102 was regularly monitored at three radio frequencies from UMRAO since early 2004, and more intensive 14.5 GHz observations were made during the \emph{RXTE} campaign.
These observations extend to early 2006.
Following the conclusion of the \emph{RXTE} campaign, optical monitoring via SMARTS continued into December, supplemented by one additional Lowell Observatory run in December.
In early 2006, UMRAO monitoring detected a large increase in the object's 14.5 GHz flux.
No optical data could be obtained at this time because the object was too close to the Sun.
However, a few \emph{RXTE} observations were obtained.
A summary of the multiwavelength data analyzed in this work is given in Table 4.

\subsection{Discussion of Multiwavelength Results}

The multiwavelength observations of CTA 102 throughout 2005 and early 2006 include many interesting events.
As displayed in Figure 7, prominent flares occurred in all three wavelength regimes.
The \emph{RXTE} campaign detected the decline following a large flaring event in which the X-ray flux changed by at least a factor of 2 ($6-7\sigma$).
The optical observations captured a few flares of $2-5\sigma$ increases in flux, with simultaneous, correlated behavior observed in the B-, V-, and R-bands.
An optical flare beginning in late September appears to decline at about the same time as the observed X-ray flux decline.
The multiband radio observations displayed in Figure 8 include all data from 2004 to early 2006.
The 14.5 GHz flare is readily apparent and is followed by an increase in the 8.0 GHz flux.
Both of these events have a significance of at least 4$\sigma$.
The flux at 4.8 GHz remains flat throughout the observations.
The X-ray observations from 2006 February are sparse, but display a sharp increase in flux of about 9$\sigma$ occurring near the time of the 14.5 GHz flare.
The X-ray flux at this time is lower than that observed in September.

The early X-ray observations indicate that the \emph{RXTE} campaign began near the peak of the X-ray flare.
Assuming the flare was symmetrical, which is usually the case in blazars, this flare almost doubled the X-ray flux and lasted about 20 days.
The optical flare was not as intensively observed, but the activity is clearly correlated across all optical bands.
The optical flux doubled over the course of about 10 days, making this flare appear similar in structure and timescale to the X-ray flare.
The observed large increase in radio flux begins simultaneously with the X-ray and optical events.
The 14.5 GHz flare, beginning at a flux around 3.5 Jy, does not double over the course of our observations.
However, our observations stop before the radio flux begins to decline, so it is not clear what was the true peak flux value.
What is clear is that the 14.5 GHz flare is much broader than the X-ray and optical flares, taking $\sim$140 days just to reach the observed peak.

To examine quantitatively whether or not the observed X-ray and optical flares were related, an Interpolated Cross-Correlation Function (ICCF) was calculated for the X-ray and R-band data to see what lag times might exist between the X-ray and R-band light curves.
\citet{wel99} has demonstrated that cross-correlation analysis, performed on data which include interpolated data to fill in gaps in observations, yields results equal to or better than those from the frequently used Discrete Correlation Function of \citet{ede88}.
Before calculating each correlation function, the flux data in the X-ray and R wavebands were normalized to the average observed flux in each waveband over the course of the 2005 \emph{RXTE} observations.
These data were then averaged every 12 hours to give a maximum of two observations per day.
Linearly interpolated data were then added to the averaged observed R-band and X-ray data, so that each data set contained an observation every 12 hours throughout the \emph{RXTE} campaign.
Our cross-correlation analysis did not reveal any significant correlations between the X-ray and optical observations.
The much lower sampling rate of the optical data compared to the X-ray data also makes it difficult to strongly argue for any correlations based on the 2005 observations, even though a visual examination of the light curves indicates a possible correlation between the drops in both optical and X-ray flux. 

\citet{tor99} observed a simultaneous 90 GHz radio and optical flare in this object, in mid-1997.
These flares were very similar in duration as well as amplitude, exhibiting fluxes in both bands increasing by a factor of 2--2.5 and lasting about 70-75 days.
\citet{tor99} note that, in 1996, a rapid optical flare was observed without any observed correlated radio activity.
As discussed in the previous section, a huge R-band flare was observed in 2004.
However, the long-term radio behavior displayed in Figure 8 shows no radio frequency flaring events around the time of the 2004 optical flare that match the optical flare's intensity or structure.
We performed ICCF analysis on the 2004--2005 14.5 GHz and R-band observations displayed in Figure 9.
The procedure was very similar to that described for the X-ray band and R-band ICCF analysis, but with a sampling rate of one point per day instead of two.
Our analysis found no significant correlations between the radio and optical activity.
These results agree with those of \citet{aom08} for the FSRQ PKS 1622--297, where no significant correlations were found between the optical and radio activity observed over several weeks.
Our results also agree with \citet{how04}, who find no significant correlations between the observed radio and optical behavior of several FSRQs over timescales of a few years.
These observations strongly suggest that different regions are dominant at different times in CTA 102, and possibly in other FSRQs.

Many recent observations of FSRQs have detected unexpected changes in color during a flaring state.
Instead of becoming bluer when in a brighter state, as is observed in BL Lac objects, many FSRQs have been observed as redder during a brighter state \citep{mil81,gu06,aom08}.
In a similar trend, \citet{rai07} observed a UV-excess in 3C 454.3 during lower flux states, indicating ``bluer-when-fainter" behavior.
Figure 10 displays the B-, V-, and R-band data collected in 2005.
A flare was observed across all three wavebands, and the flare structure is very similar in all three wavebands.
Figure 11 displays the difference in flare amplitude across wavelength as measured by changes in flux and brightness. 
During the flare, the R-band flux increases more than the V-band flux, which in turn increases more than the B-band flux.
In our 2005 observations of CTA 102, it appears that this object appears somewhat redder as the object brightens.
Our 2004 campaigned observed CTA 102 in a particularly bright state, but optical observations were only obtained in the R-band at this time, thus prohibiting any color analysis on this data. 
More observations of CTA 102 during various flux states are needed to confirm this trend, preferably including more extensive coverage in the infrared.

\subsection{Broadband Spectral Analysis}

To examine how the spectral shape, or SED, changed over time, the 2005 campaign on CTA 102 was first split into four bins of one to a few days in size.
The bins were defined based on times separating increases or decreases in the X-ray and optical flux.
Figure 12 displays the \emph{RXTE} data from each campaign with the bins, X1--X4, marked.
This figure also displays the simultaneous optical data, again with bins X1--X4 labeled in time.
The sampling of the radio data was not always dense enough at 4.8 and 8.0 GHz to provide simultaneous data for each time bin.
However, the flux changes occur at slow enough rates at these frequencies so that a given time bin's data could be linearly interpolated from real 4.8 and 8.0 GHz data.
The resulting SEDs for CTA 102 are plotted in Figure 13. 
The optical differential magnitudes were converted to apparent magnitudes, then to fluxes using the method of \citet{cox01}.
An average frequency of $1.5\times10^{18}$ Hz, corresponding to about 6 keV, was used for the \emph{RXTE} data.

The broad radio to X-ray spectral shape of CTA 102 suggests that the X-ray emission could be due to both the Synchrotron Self-Compton (SSC) and external IC processes.
The broadband spectral shape does not significantly change over the course of the \emph{RXTE} campaign.
The results in Figure 13 are very similar to an SED published in \citet{tav00} using non-simultaneous data on this object.
They successfully fit the SED with a homogenous external Compton (EC) model, with the X-ray emission produced by relativistic electrons upscattering photons from the accretion disk or the broad-line region (BLR) \citep[][ and references therein]{tav00}.
\citet{tav00} also suggest that both the SSC and EC models could contribute to \emph{RXTE} regime emission.
For this to be the case, the X-ray emission could at times be due to the jet synchrotron photons being upscattered by the relativistic electrons in the jet to produce the IC X-ray emission.
In results of long-term monitoring of 3C 273 using multiwavelength observations, the broadband spectral analysis suggests that both SSC and EC processes may be present in this object at different times \citep{sol08}.

\section{Conclusions}

Our analysis of the optical microvariability of CTA 102 suggests that the character of the microvariability changes in response to the overall optical state.
In higher flux states, microvariability events feature more rapid changes in flux compared to events observed at lower flux states (see Figure 6).
Our results appear to contradict those of \citet{sta04}.
The results of \citet{sta04} suggest that the mechanisms behind optical microvariability and long-term variability are different, and thus should not appear to exhibit correlated behavior.
Investigating correlations between flux state and the character of microvariability is an important step toward understanding the underlying physics behind blazar microvariability.
\citet{how04} suggest that microvariability can be produced by a variety of mechanisms, including instabilities in the particle acceleration mechanism, variations in the supply of electrons, or small inhomogeneities in the object's magnetic fields or the ambient medium.
Others suggest that accretion disk instabilities and fluctuations can cause microvariations \citep{man93,sta04,gup05}.
However, \citet{how04} argue that accretion disk phenomena are not responsible for microvariations because such phenomena would be washed out by the highly beamed jet emission.
Additional campaigns are needed on CTA 102 and other blazars to more comprehensively examine possible relationships between the optical flux state and optical microvariability activity in these objects.

Significant optical microvariability has been observed in this object, but not in conjunction with remarkable radio or X-ray activity. 
In 2004, the most dramatic microvariability observed occurred simultaneously with no flaring activity in the radio regime. 
In 2005, significant X-ray, optical, and radio flaring was observed, but with no significant optical microvariability events. 
This could indicate that the processes causing optical microvariability are not related to larger scale jet events affecting the synchrotron emission in this object, which would likely cause correlated optical and radio events. 
The cause of the lack of microvariability with X-ray activity needs further investigation in future campaigns. 
One possible explanation for these missing correlations would be that the optical microvariability is at times not related to jet processes, thus arguing in favor of accretion disk activity causing optical microvariability.
However, the 2005 campaign was the only campaign where X-ray, optical, and radio data were all available. 
More broadband campaigns with more even sampling across regimes are needed to properly investigate the microvariability-jet and microvariability-accretion disk connections.

Investigations of the optical emission show that CTA 102 appears redder when in a brighter flux state.
\citet{gu06} suggest that this behavior could be a result of the ``blue bump" observed in many FSRQs, thought to be thermal accretion disk radiation.
Depending on how large this thermal contribution is, it could be significant in lower optical flux states and overwhelmed by the jet emission in higher states, thus making the FSRQ appear redder when brighter.
Recent campaigns on similar objects have found significant blue bump contributions to the observed spectra, e.g., \citet{kat08,rai08}.
Future campaigns on FSRQs should include UV observations to better show how strong the blue bump component is compared to the continuum.

Broadband spectral analysis of CTA 102 suggests that the X-ray emission could be due to either the synchrotron or the IC process. 
Multiyear \emph{Swift}/BAT transient monitor results provided by the \emph{Swift}/BAT team\footnote{http://swift.gsfc.nasa.gov/docs/swift/results/transients/index.html, CTA 102 is listed as 4C 11.69.} indicate that CTA 102 was in a relatively low X-ray flux state at the time of our 2005 campaign. 
Additional campaigns at varying levels of X-ray flux are needed to better determine the source of the X-ray emission in this object.
Such campaigns would greatly benefit from longer temporal coverage in order to find clearer correlations between the synchrotron and X-ray behavior.
It is also important to include UV and $\gamma$-ray observations in future campaigns in order to better investigate the high-energy synchrotron tail and IC peak of CTA 102 and similar objects.

\section{Acknowledgments}

A.O.M., H.R.M., K.M., and W.T.R. were supported in part for this work by the Program for Extragalactic Astronomy's Research Program Enhancement funds from Georgia State University. 
The \emph{RXTE} observations were supported by NASA grants including NAG5-13733.
A.O.M., H.R.M., K.M., and W.T.R. thank Lowell Observatory and Boston University for generous allocations of observing time on the Perkins telescope with the Loral and PRISM cameras.
They also thank Lowell Observatory for generous allocations of observing time on the Hall Telescope.
The authors thank Todd Henry and the SMARTS consortium for contributing optical observations to this work.
The UMRAO facility is partially supported by a series of grants from the NSF, most recently NSF-0607523, and by the University of Michigan.

\begin{deluxetable}{ccc}
\tablewidth{0pt}
\tablecaption{Lowell Observatory and SMARTS Detector Information}
\tablehead{
\colhead{Telescope} & \colhead{CCD} & \colhead{Chip Size}\\
& & \colhead{(pixels)}
}
\startdata
Lowell Hall 1.1m/Perkins 1.8 m &USNO &800$\times$800\\
&SITe &2048$\times$2048\\
&LORAL &2048$\times$2048\\
Lowell Perkins 1.8 m &PRISM &2048$\times$2048\\
CTIO 1.3 m &FAIRCHILD 447 &2048$\times$2048\\
\enddata
\end{deluxetable}

\begin{deluxetable}{cccc}
\tablewidth{0pt}
\tablecaption{CTA 102 Comparison Star Apparent Magnitudes.}
\tablehead{
\colhead{Check ID} & \colhead{B mag} & \colhead{V mag} & \colhead{R mag} 
}
\startdata

1\tablenotemark{a} &14.77(0.04) &13.98(0.03) &13.56(0.04)\\
2\tablenotemark{a} &16.17(0.04) &14.88(0.03) &14.07(0.07)\\
3\tablenotemark{b} & & &15.91(0.02)\\
z\tablenotemark{c} & &15.38(0.05) &14.70(0.03)\\
\enddata
\tablenotetext{a}{\citet{rai98}.}
\tablenotetext{b}{Calibrated from 2004 October 4 and 7 observations of comparison stars 1, 2, and z.}
\tablenotetext{c}{\citet{bec97}.}
\end{deluxetable}

\begin{deluxetable}{cccccc}
\tablewidth{0pt}
\tablecaption{Summary of CTA 102 Microvariability Events.}
\tablehead{
\colhead{UT Date} & \colhead{No. of Observations} & \colhead{Avg. R Mag.} & \colhead{Event} & \colhead{$\sigma$} & \colhead{Rate}\\
& & & & & (magnitudes/day)
}
\startdata
2000 Jun 07 & 51 & 16.46 $\pm$ 0.01 & 0 & ... & ...\\
2002 Sep 01 & 27 & 15.42 $\pm$ 0.01 & 1 & 4.29 & 0.43\\
2002 Sep 02 & 30 & 15.39 $\pm$ 0.01 & 1 & 10.0 & 0.41\\
2002 Sep 03 & 19 & 15.33 $\pm$ 0.01 & 1 & 13.3 & 0.50\\
2002 Sep 04 & 16 & 15.37 $\pm$ 0.01 & 0 & ... & ...\\
2004 Sep 21 & 102 & 14.77 $\pm$ 0.02 & 1 & 12.2 & --6.08\\
2004 Sep 22 & 280 & 14.78 $\pm$ 0.02 & 1 & 2.27 & --1.09\\
2004 Sep 23 & 243 & 14.70 $\pm$ 0.02 & 0 & ... & ...\\
2004 Sep 24 & 36 & 14.95 $\pm$ 0.02 & 0 & ... & ...\\
2004 Oct 04 & 198 & 14.53 $\pm$ 0.02 & 1 & 8.00 & 2.00\\
& & & 2 & 13.0 & --6.50\\
& & & 3 & 17.0 & 8.50\\
2004 Oct 05 & 279 & 15.10 $\pm$ 0.02 & 1 & 75.0 & 1.18\\
2004 Oct 06 & 179 & 15.42 $\pm$ 0.02 & 1 & 10.7 & 0.63\\
2004 Oct 07 & 239 & 15.61 $\pm$ 0.02 & 1 & 30.0 & --0.40\\
2004 Oct 08 & 240 & 15.53 $\pm$ 0.02 & 0 & ... & ...\\
2004 Oct 09 & 245 & 15.65 $\pm$ 0.02 & 1 & 7.00 & 0.39\\
& & & 2 & 14.0 & --0.41\\
2005 Oct 10 & 67 & 15.72 $\pm$ 0.02 & 0 & ... & ...\\
2005 Oct 11 & 84 & 15.96 $\pm$ 0.02 & 0 & ... & ...\\
\enddata
\end{deluxetable}

\begin{deluxetable}{cccc}
\tablewidth{0pt}
\tablecaption{Summary of 2005 Multiwavelength Observations.}
\tablehead{
\colhead{Regime} & \colhead{Observatory} & \colhead{Dates of Observations} & \colhead{No. of Observations}
}
\startdata
$2-10$ keV & RXTE & 2005 Sep 25$ - $Oct 15 & 35 \\
& & 2006 Feb $2 - 8$ & 3 \\
B-band & Lowell & 2005 Oct $10 - 11$ & 9 \\
& SMARTS & 2005 Sep$ - $Nov & 10 \\
V-band & Lowell & 2005 Oct $10 - 13$ & 18 \\
& SMARTS & 2005 Sep$ - $Dec & 10 \\
R-band & Lowell & 2005 May $8 - 11$ & 7 \\
& & 2005 June $26 - 28$ & 6 \\
& & 2005 Sep $12 - 14$ & 9 \\
& & 2005 Oct $10 - 13$ & 161 \\
& & 2005 Dec $10 - 13$ & 6 \\
& Foggy Bottom & 2005 May$ - $Sep & 19 \\
& SMARTS & 2005 Sep$ - $Dec & 14 \\
14.5 GHz & UMRAO & 2004 Jan$ - $2006 Jan & 88 \\
8.0 GHz & UMRAO & 2004 Jan$ - $2006 Jan & 42 \\
4.8 GHz & UMRAO & 2004 Apr$ - $2006 Feb & 29 \\
\enddata
\end{deluxetable}

\begin{figure}
\includegraphics{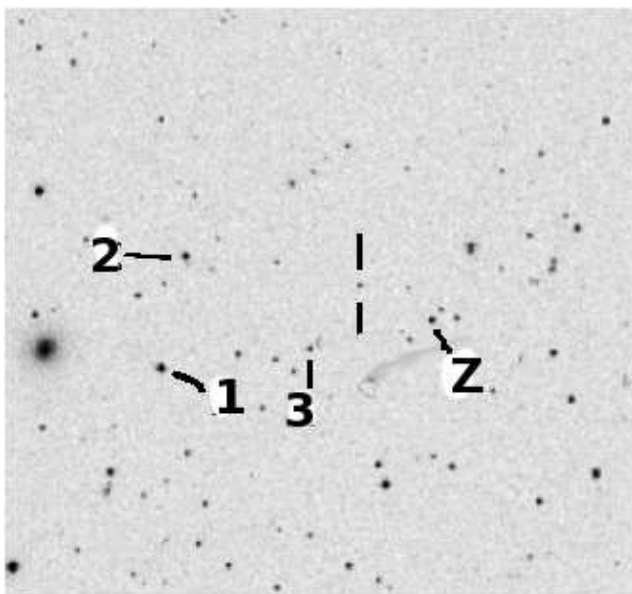}
\figcaption{Optical field of view of CTA 102, 11 arcmin by 10 arcmin, oriented such that up is north and left is east.
The object's location is indicated by two dashes.
Calibrated comparison stars 1, 2, 3, and z are labeled.
In Figures 3--5, "check1" refers to comparison star 1 above, "check2" to comparison star 2, etc. }
\end{figure}

\begin{figure}
\includegraphics[angle=270, scale=0.65]{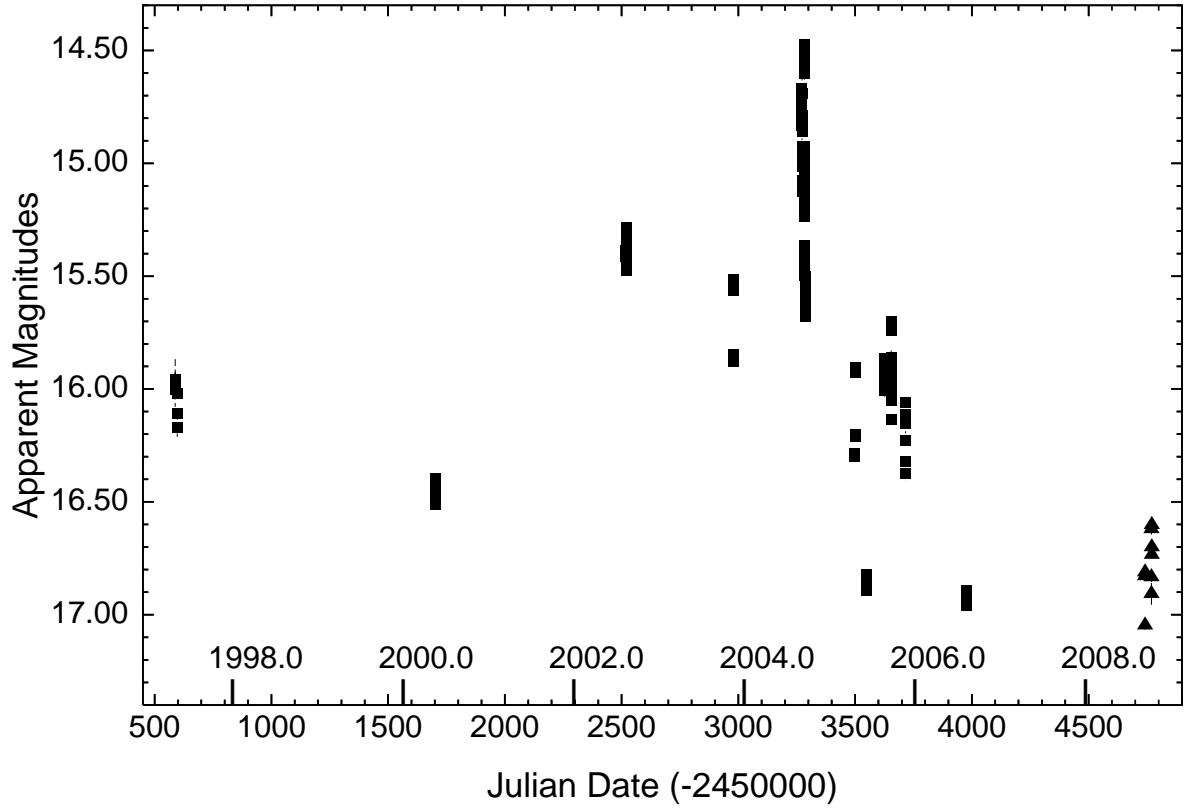}
\figcaption{Long-term R-band activity observed in CTA 102 from Lowell Observatory (squares, 2006 and earlier data) and Evelyn L. Egan Observatory (triangles, 2008 data).}
\end{figure}

\begin{figure}
\centering
\includegraphics[angle=270, scale=0.32]{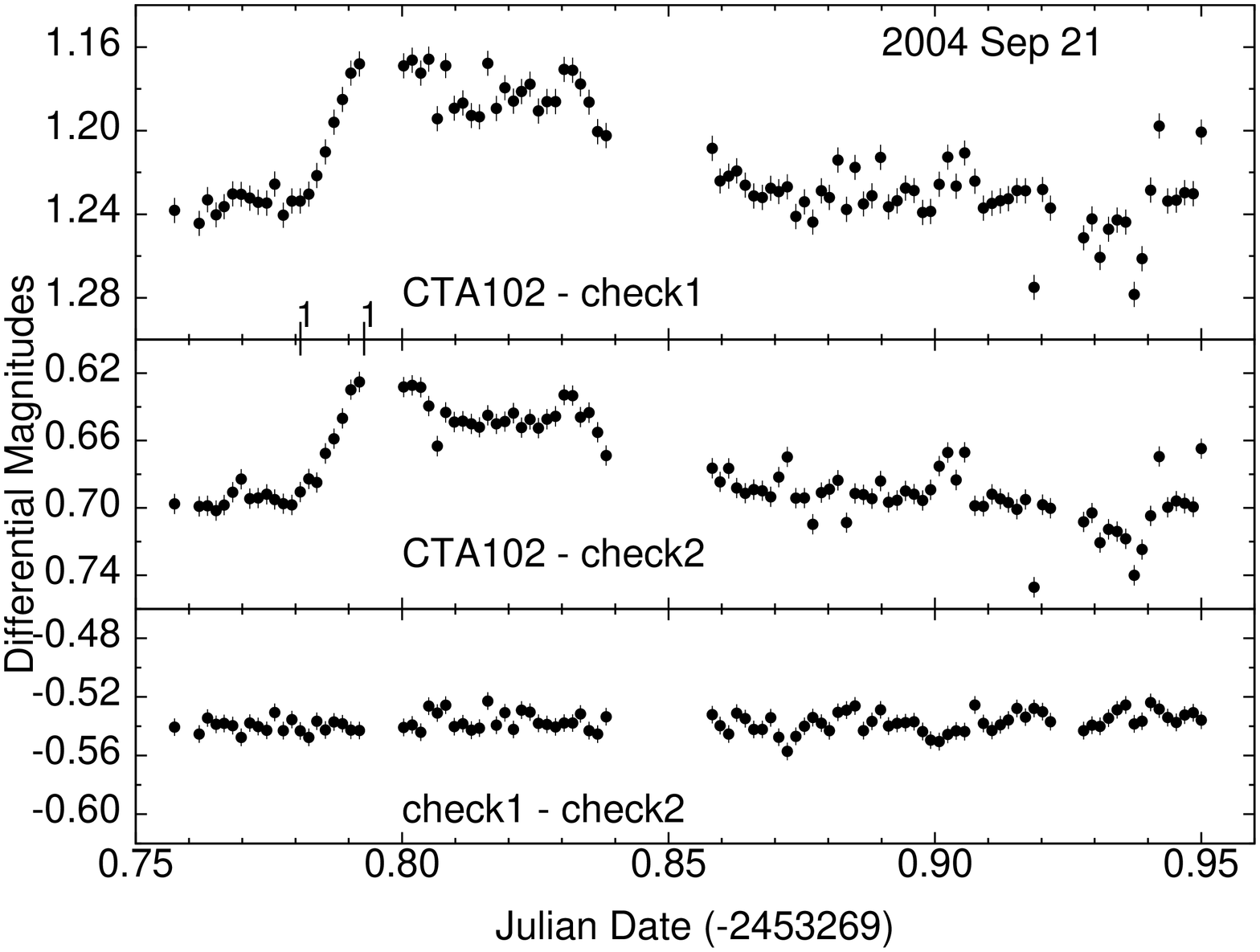}
\includegraphics[angle=270, scale=0.32]{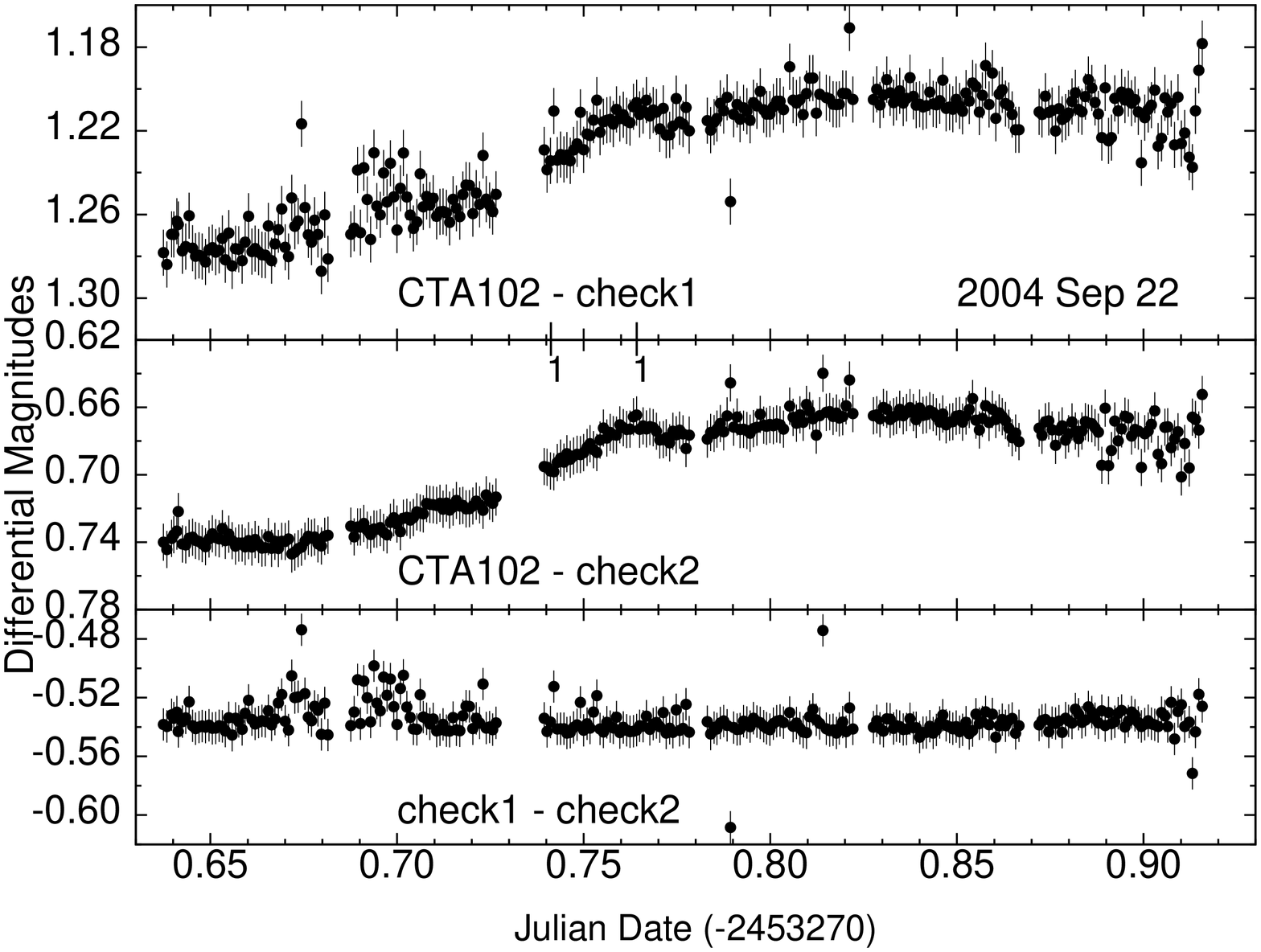}\\
\includegraphics[angle=270, scale=0.32]{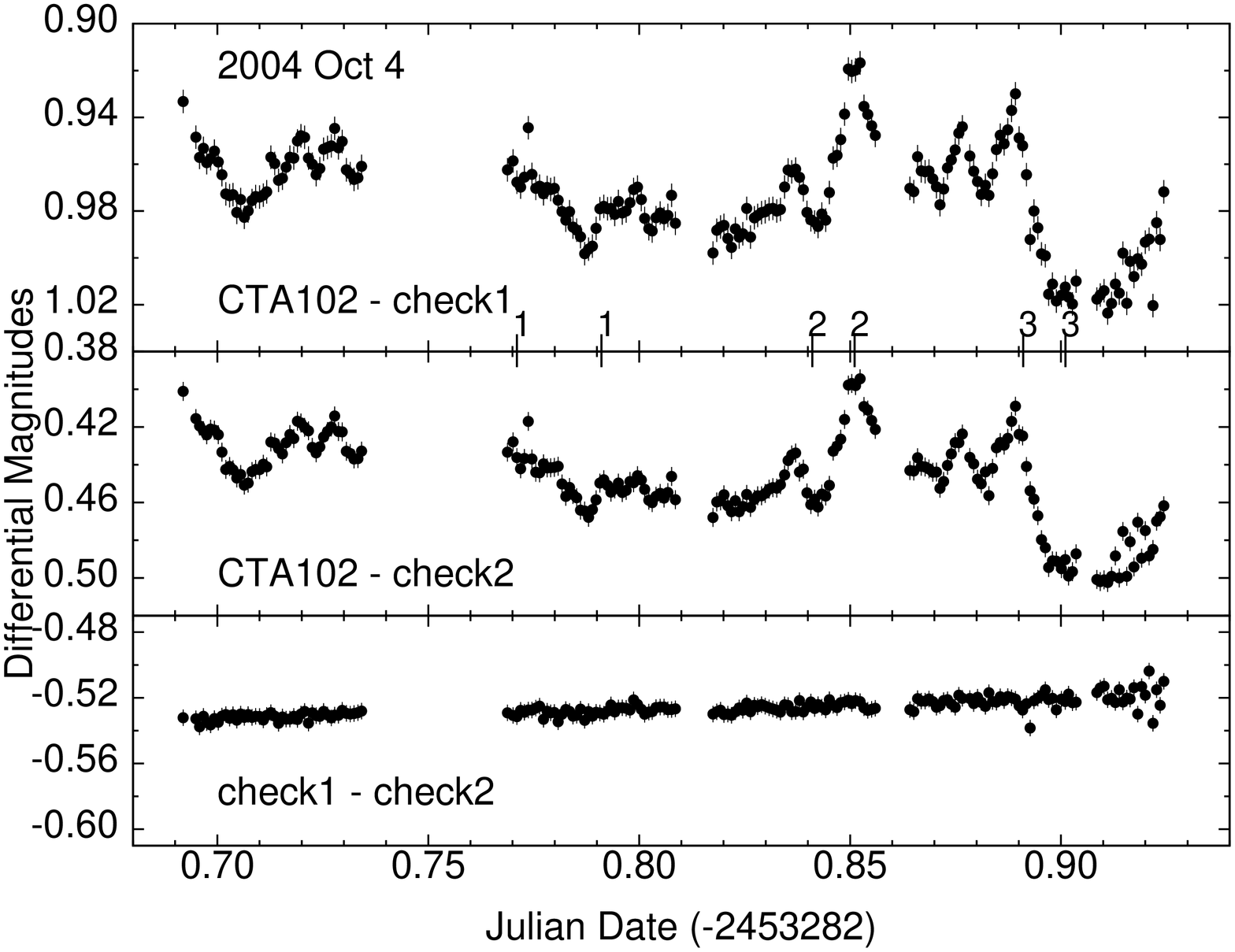}
\includegraphics[angle=270, scale=0.32]{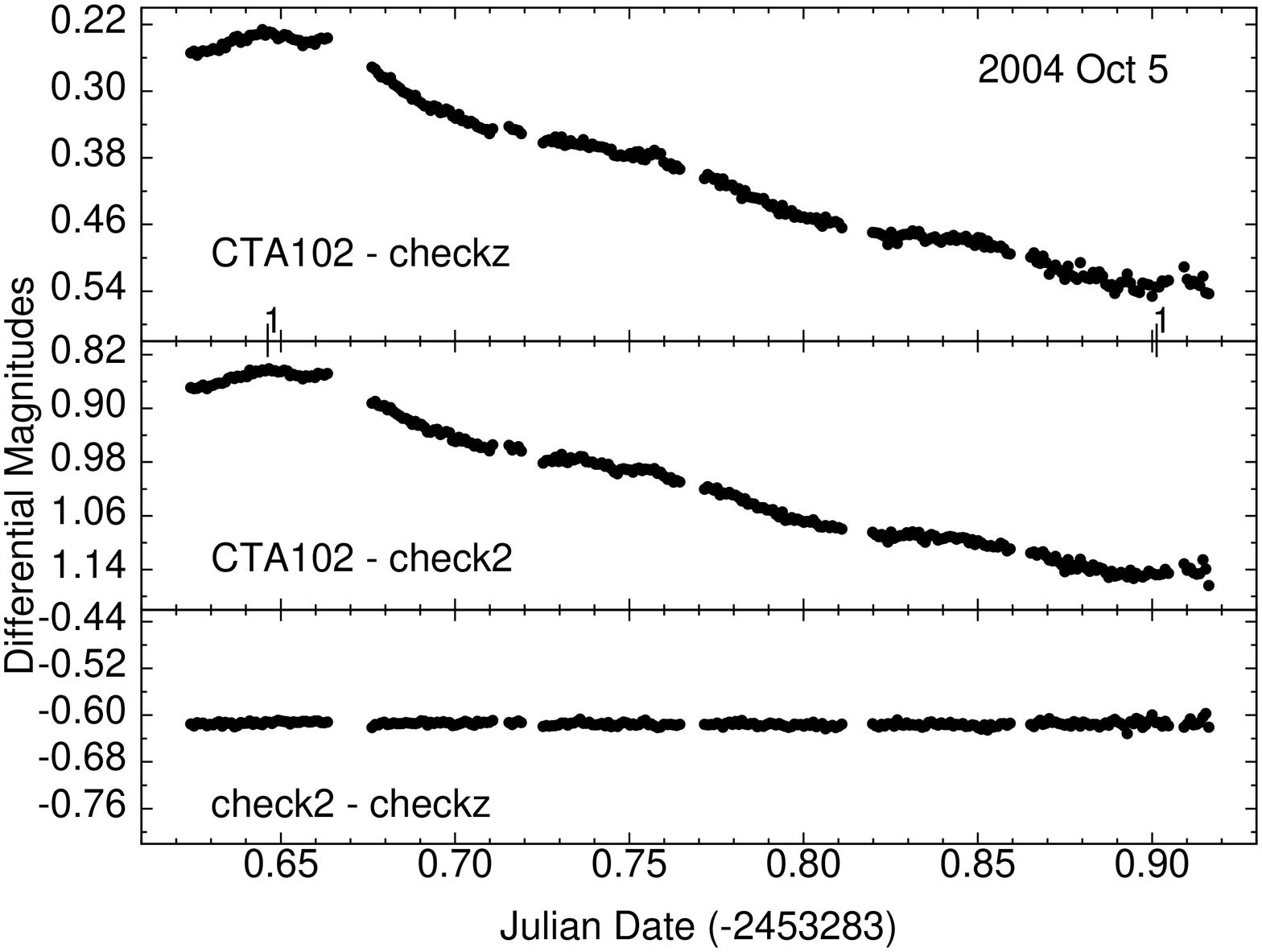}\\
\caption[microvara]{Microvariability events observed in 2004 when CTA 102 was in the brightest observed optical state from 1997 to 2006. 
The beginning and end of each event described in Table 3 are labeled on the time axes. }\label{microvara}
\end{figure}

\begin{figure}
\centering
\includegraphics[angle=270, scale=0.32]{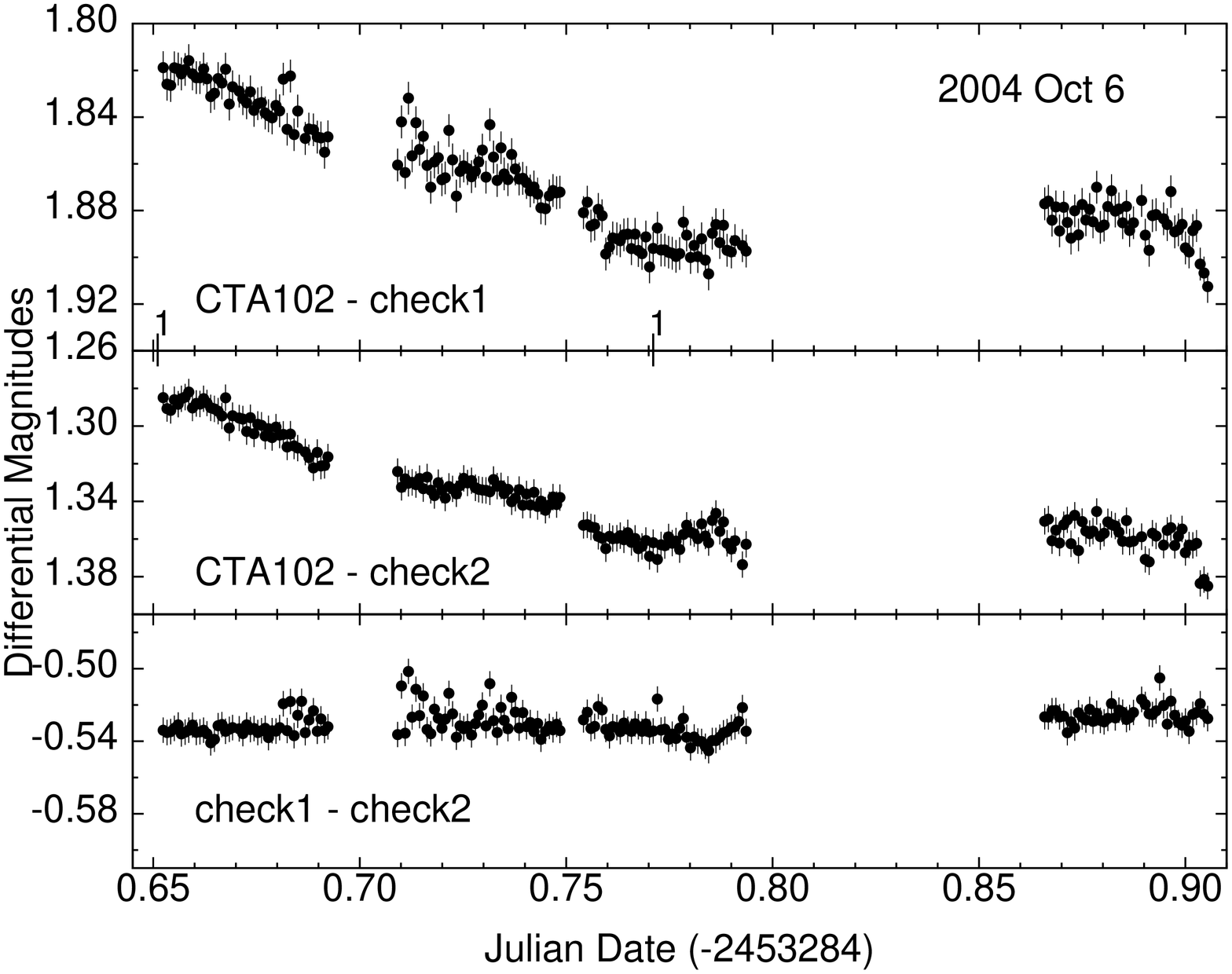}
\includegraphics[angle=270, scale=0.32]{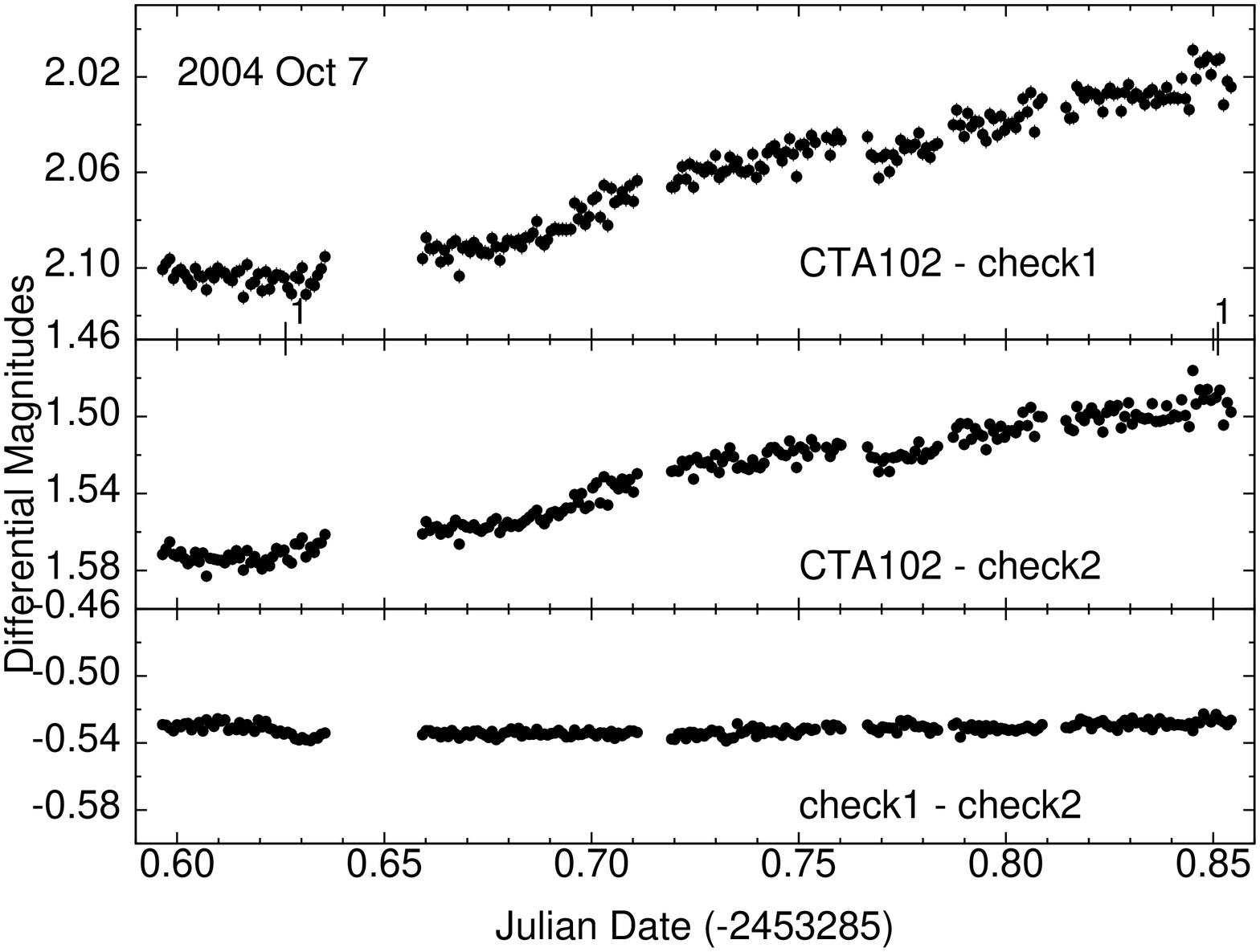}\\
\includegraphics[angle=270, scale=0.32]{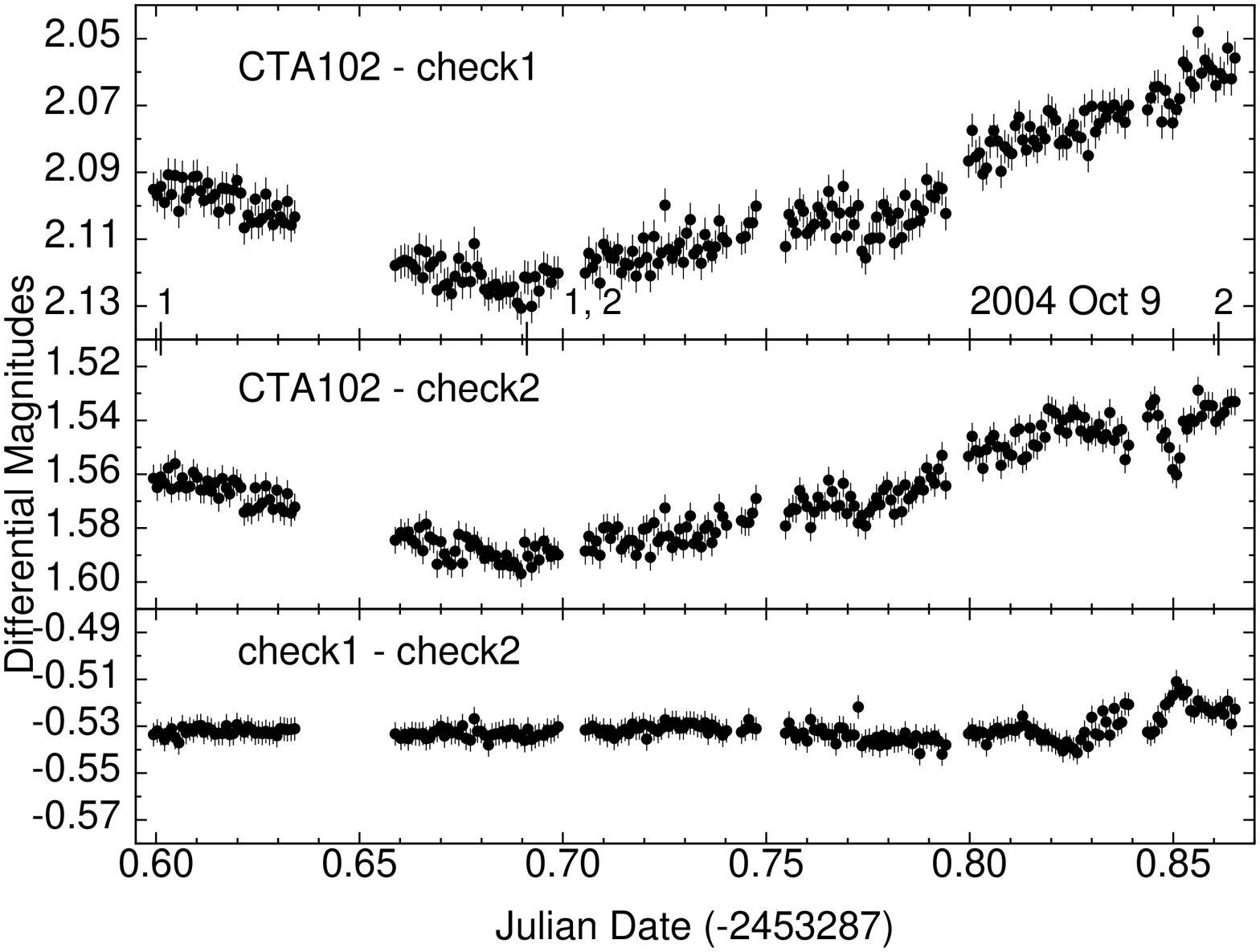}\\
\caption[microvarb]{Microvariability events observed in 2004 at the end of the large optical outburst. 
The beginning and end of each event described in Table 3 are labeled on the time axes. }\label{microvarb}
\end{figure}

\begin{figure}
\centering
\includegraphics[angle=270, scale=0.32]{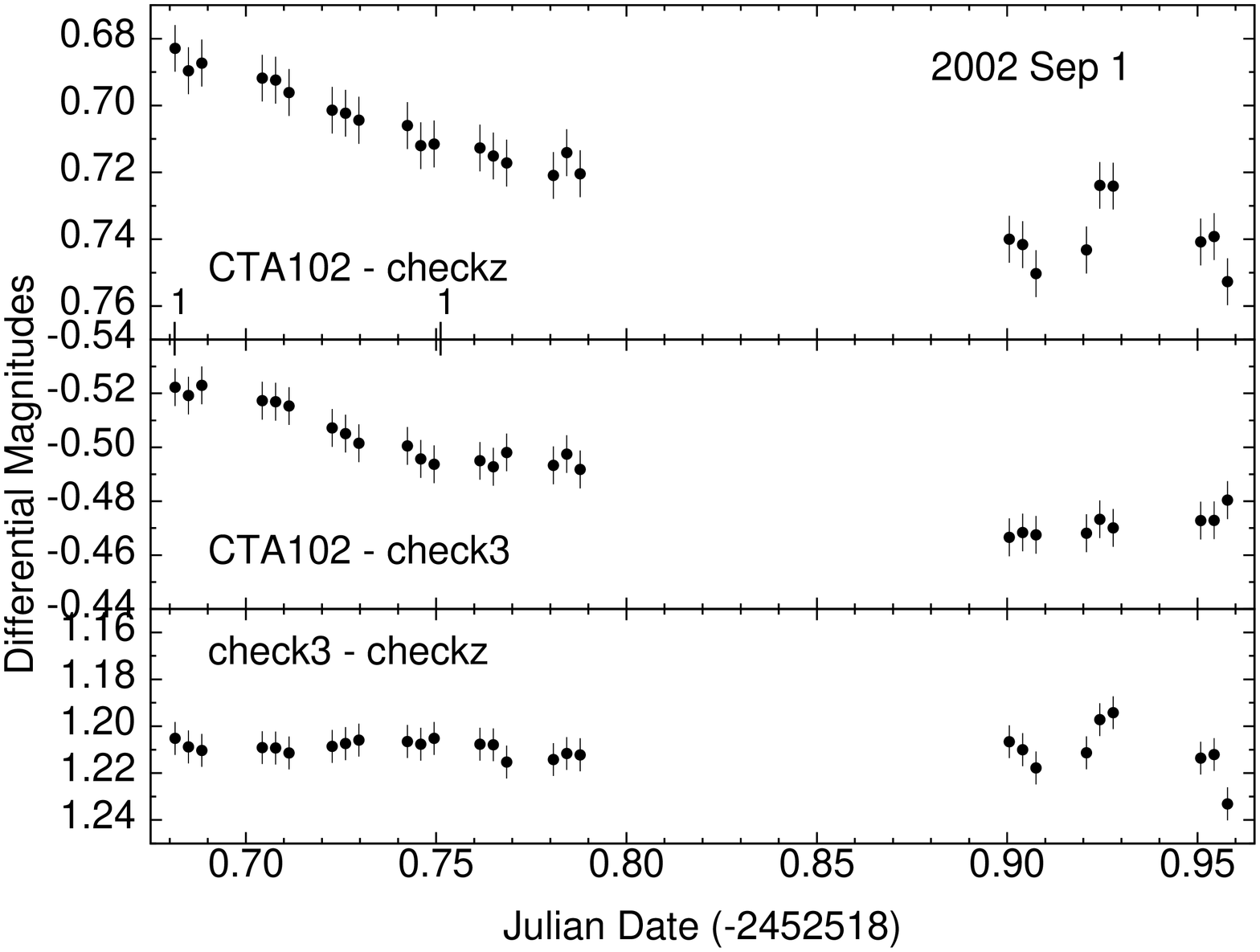}
\includegraphics[angle=270, scale=0.32]{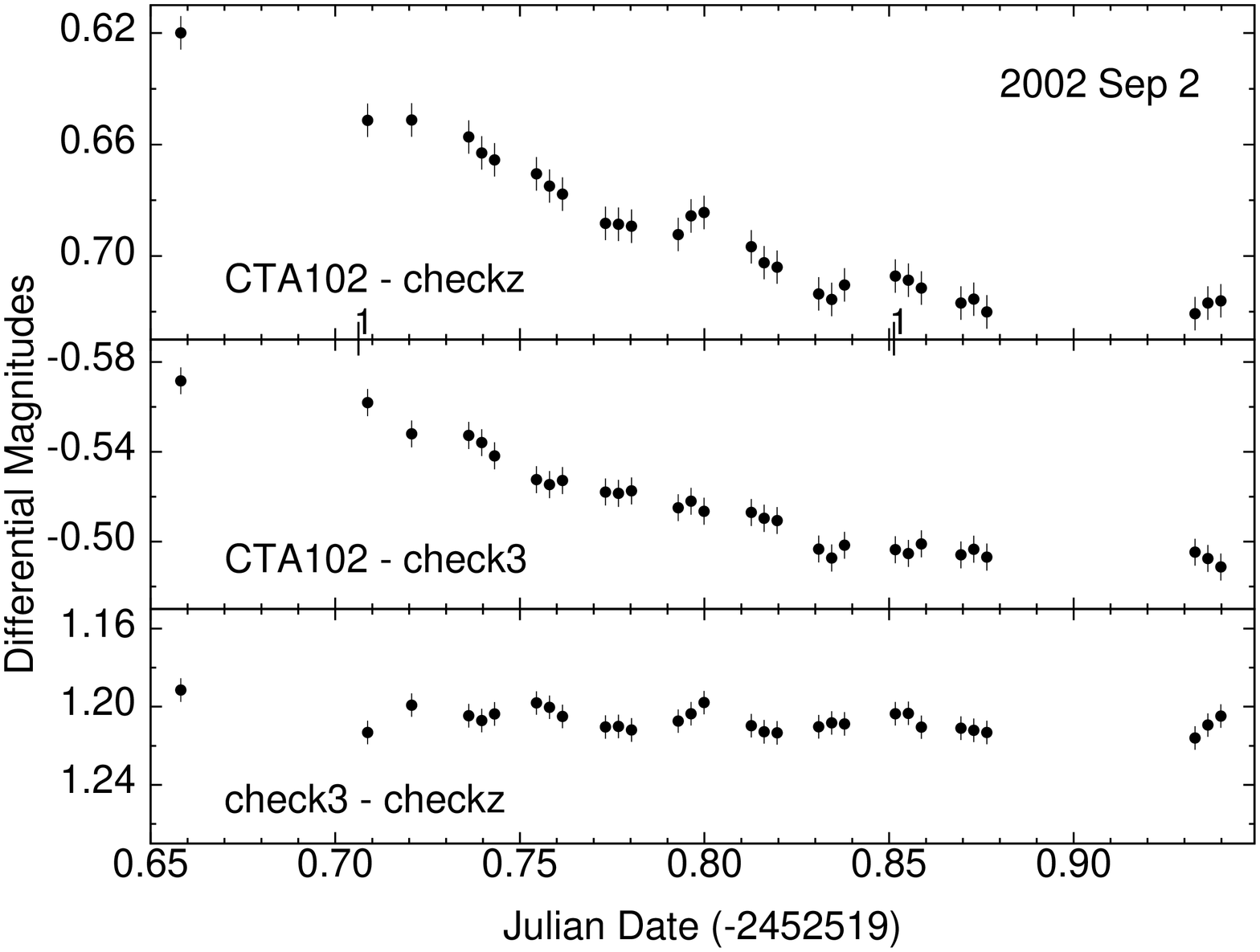}\\
\includegraphics[angle=270, scale=0.32]{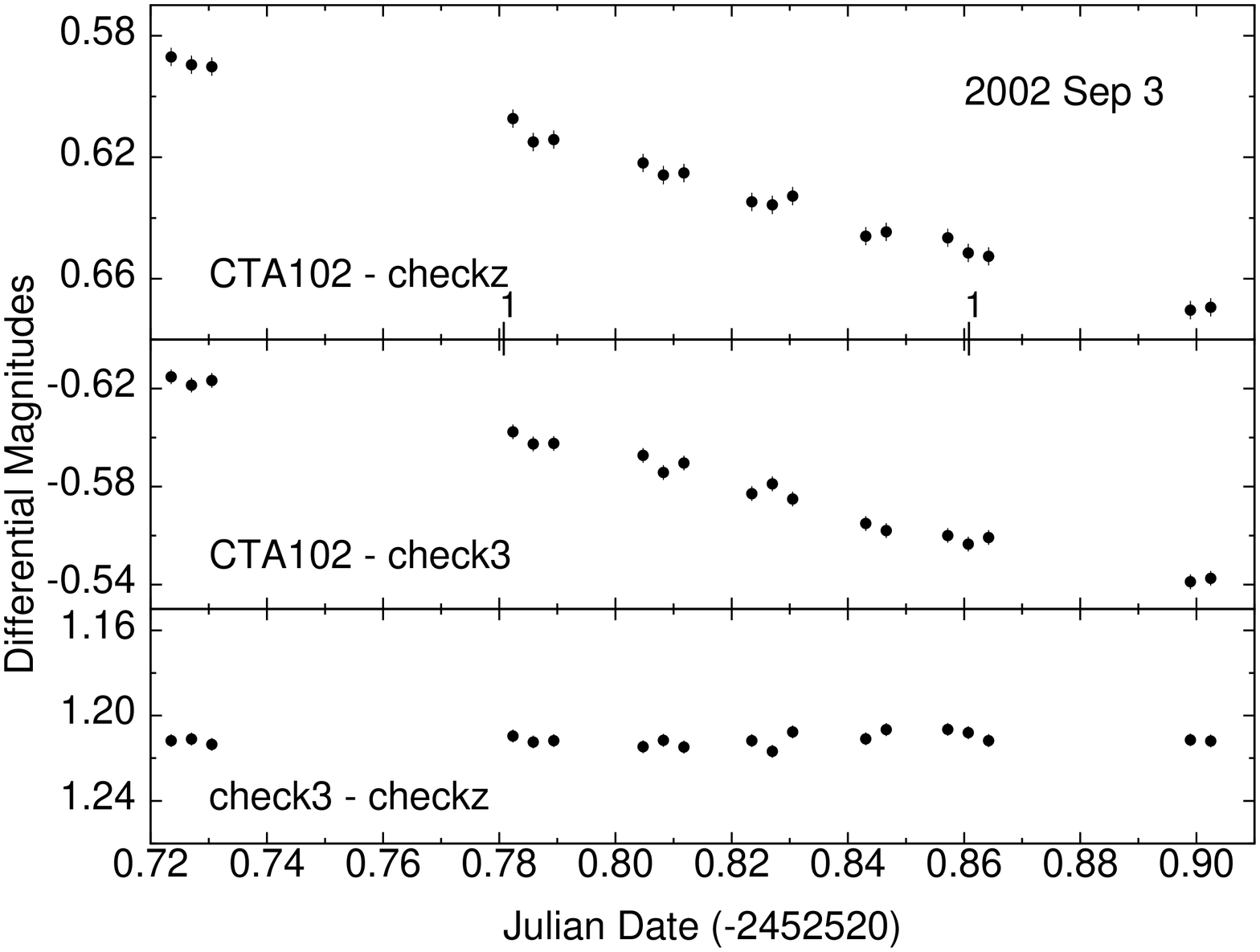}\\
\caption[microvarc]{Microvariability events observed in 2002 when CTA 102 was in an elevated flux state. 
The beginning and end of each event described in Table 3 are labeled on the time axes. }\label{microvarc}
\end{figure}

\begin{figure}
\includegraphics[angle=270, scale=0.65]{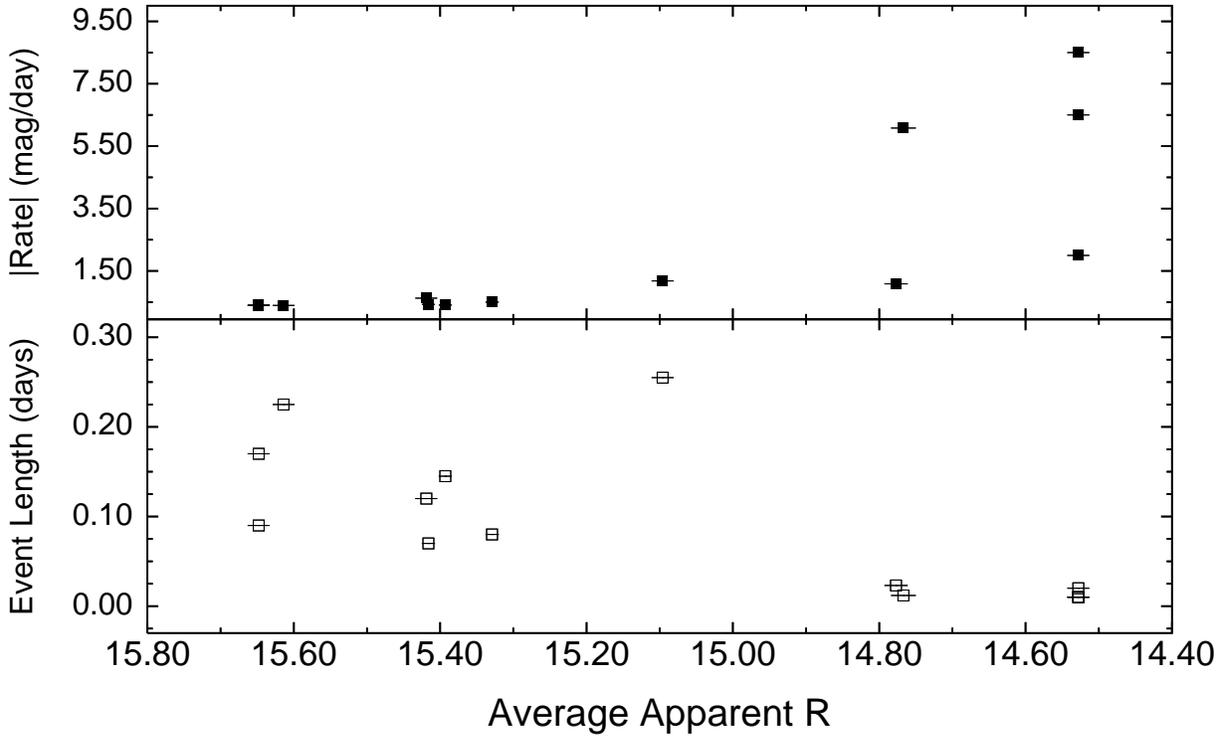}
\figcaption{Comparisons of flaring event duration (days) and rate of brightness change (mag/day) with the average observed brightness on the night of the flaring event.}
\end{figure}

\begin{figure}
\includegraphics[angle=270, scale=0.65]{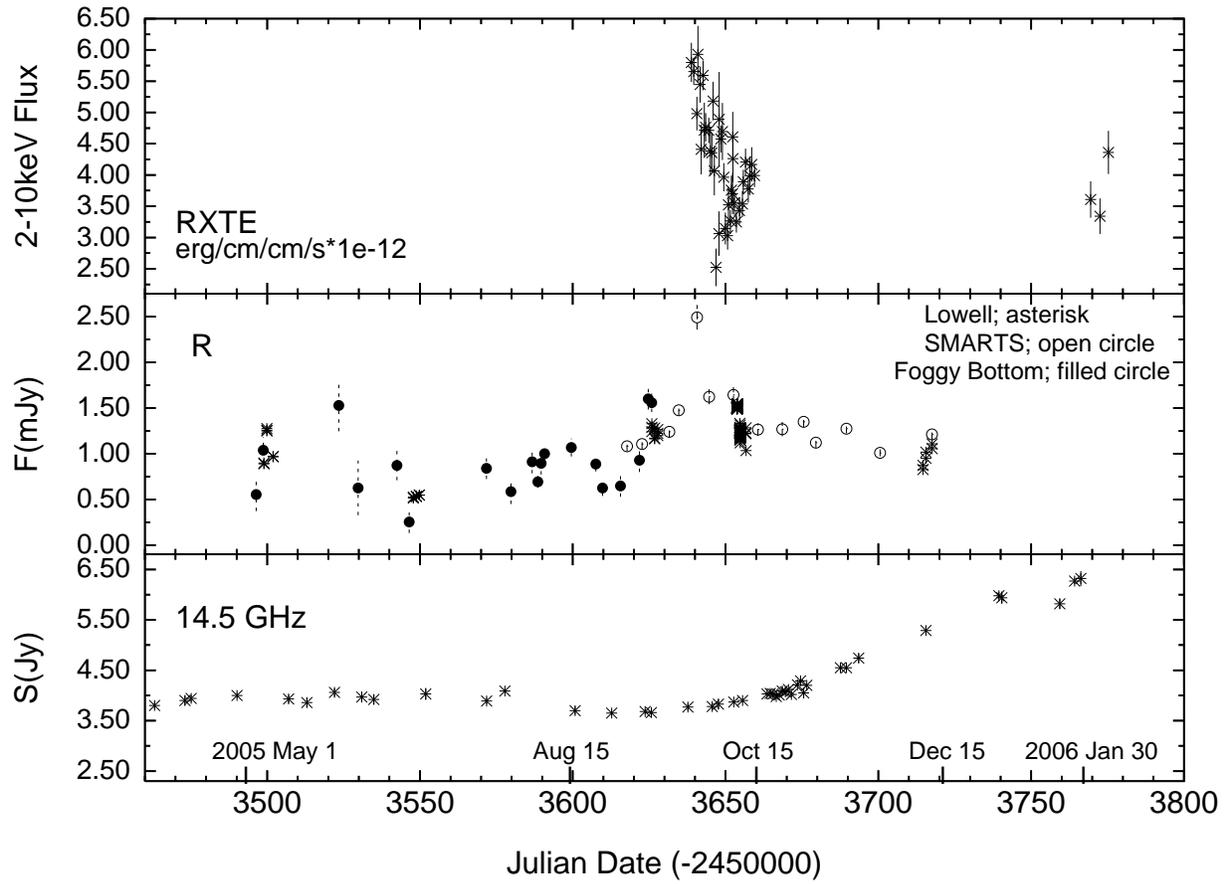}
\figcaption{Simultaneous multiwavelength observations of CTA 102 from the 2005 campaign.}
\end{figure}

\begin{figure}
\includegraphics[angle=270, scale=0.65]{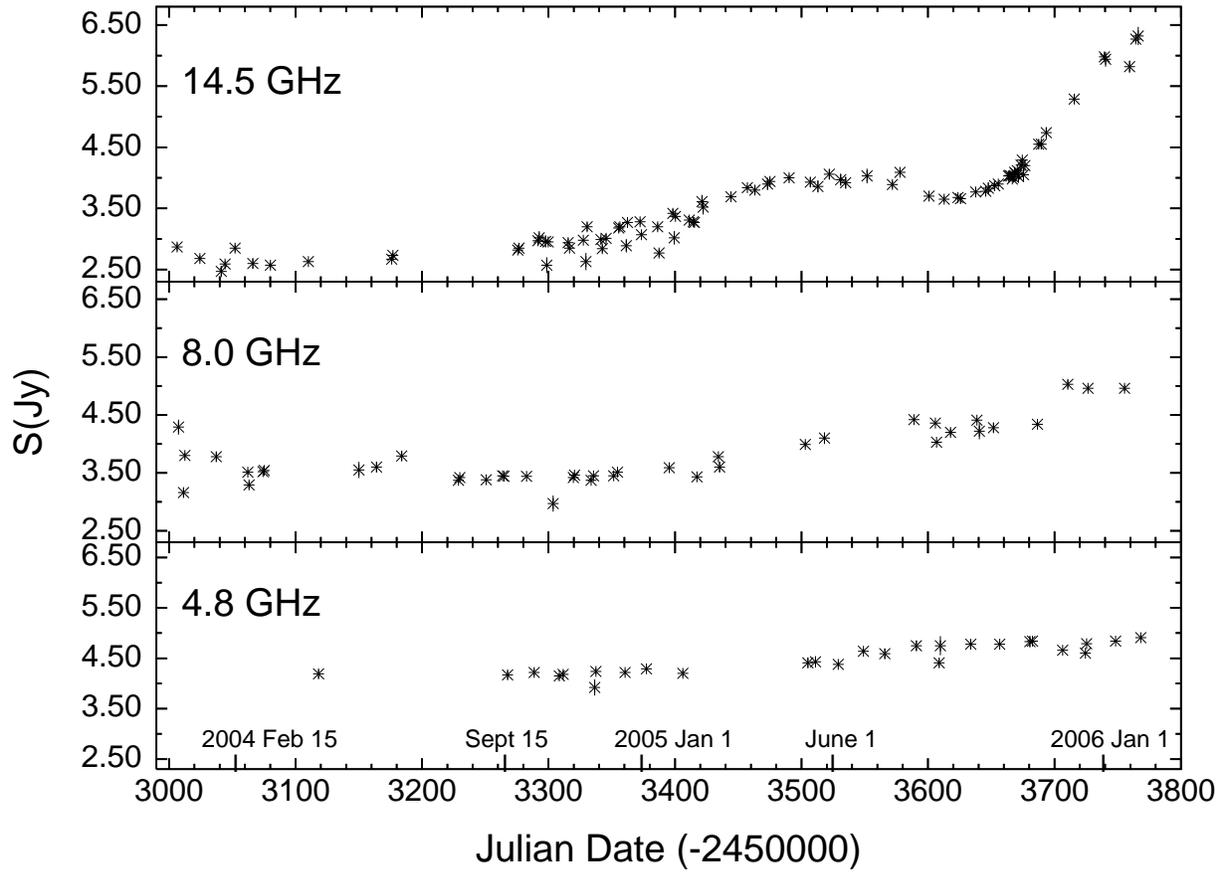}
\figcaption{Long-term radio activity observed in CTA 102 from UMRAO.}
\end{figure}

\begin{figure}
\includegraphics[angle=270, scale=0.65]{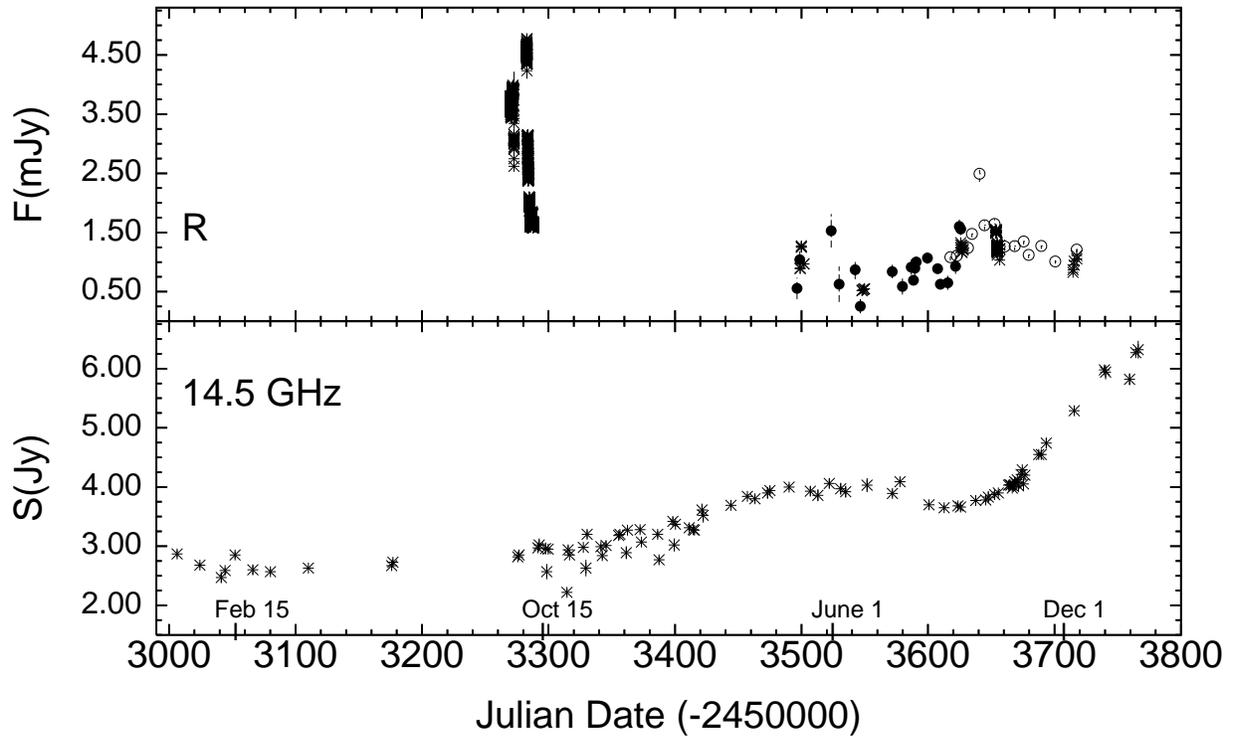}
\figcaption{Simultaneous R-band and 14.5 GHz observations of CTA 102 from 2004 and 2005. 
See Figure 7 for the key to the plotted optical data symbols. }
\end{figure}

\begin{figure}
\includegraphics[angle=270, scale=0.40]{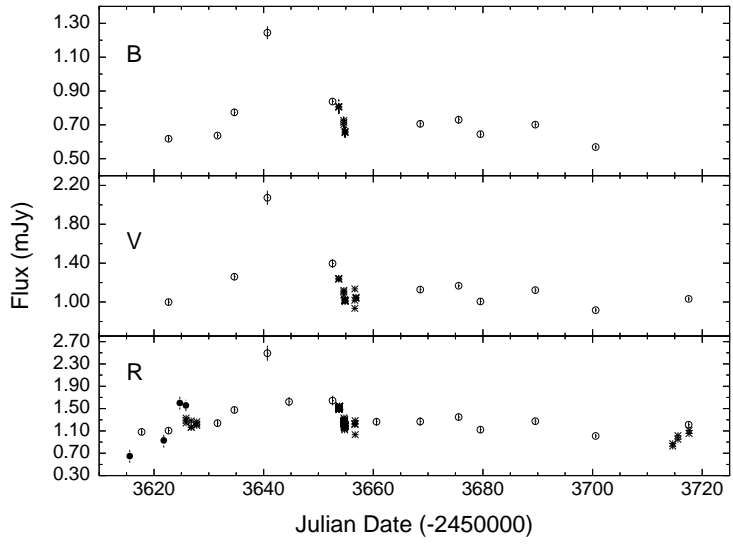}
\figcaption{Optical color behavior of CTA 102 observed in 2005.
This figure displays the similarities in structure and timescale across the B-, V-, and R-bands.
See Figure 7 for the key to the plotted optical data symbols. }
\end{figure}

\begin{figure}
\includegraphics[angle=270, scale=0.40]{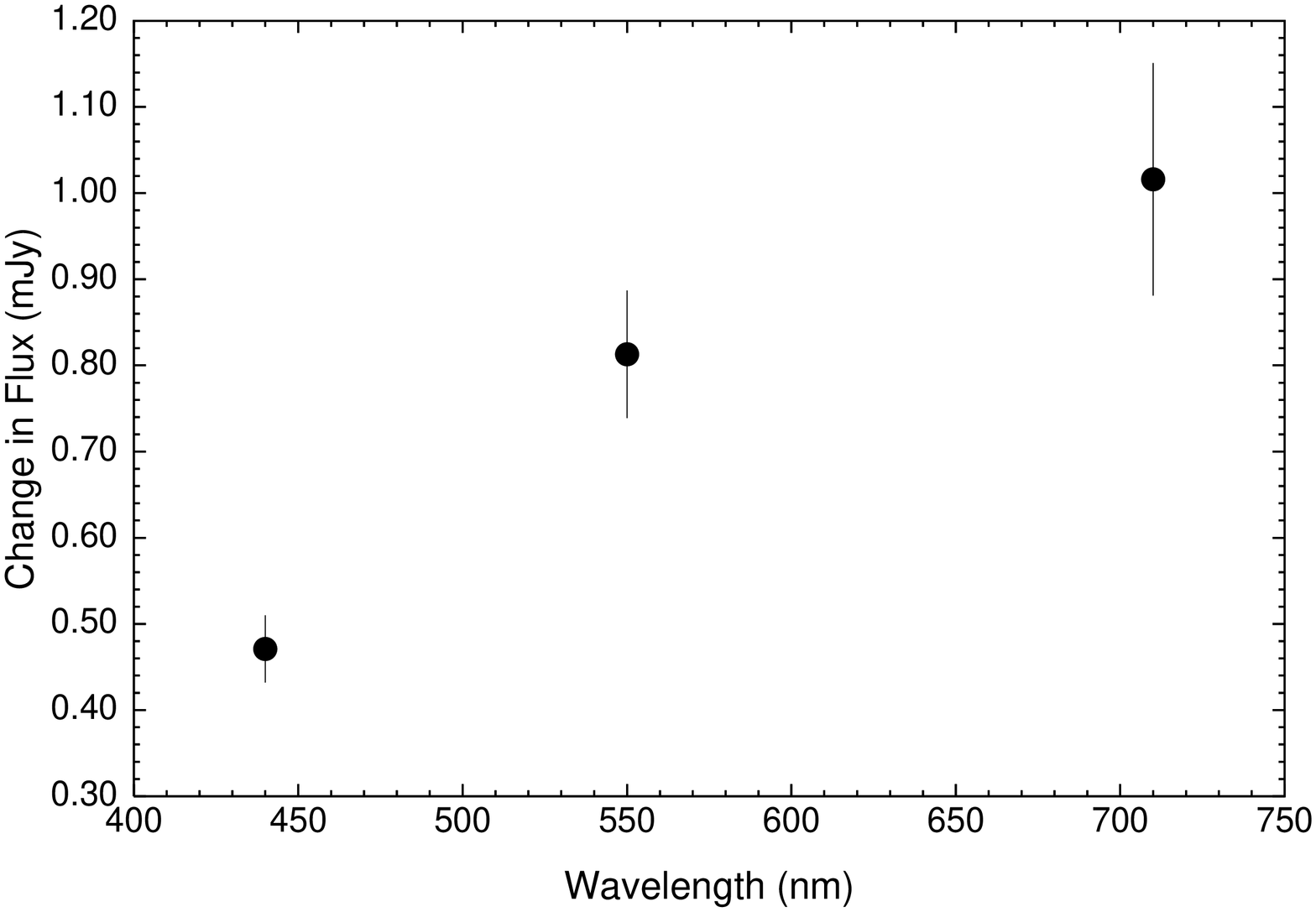}\\
\includegraphics[angle=270, scale=0.40]{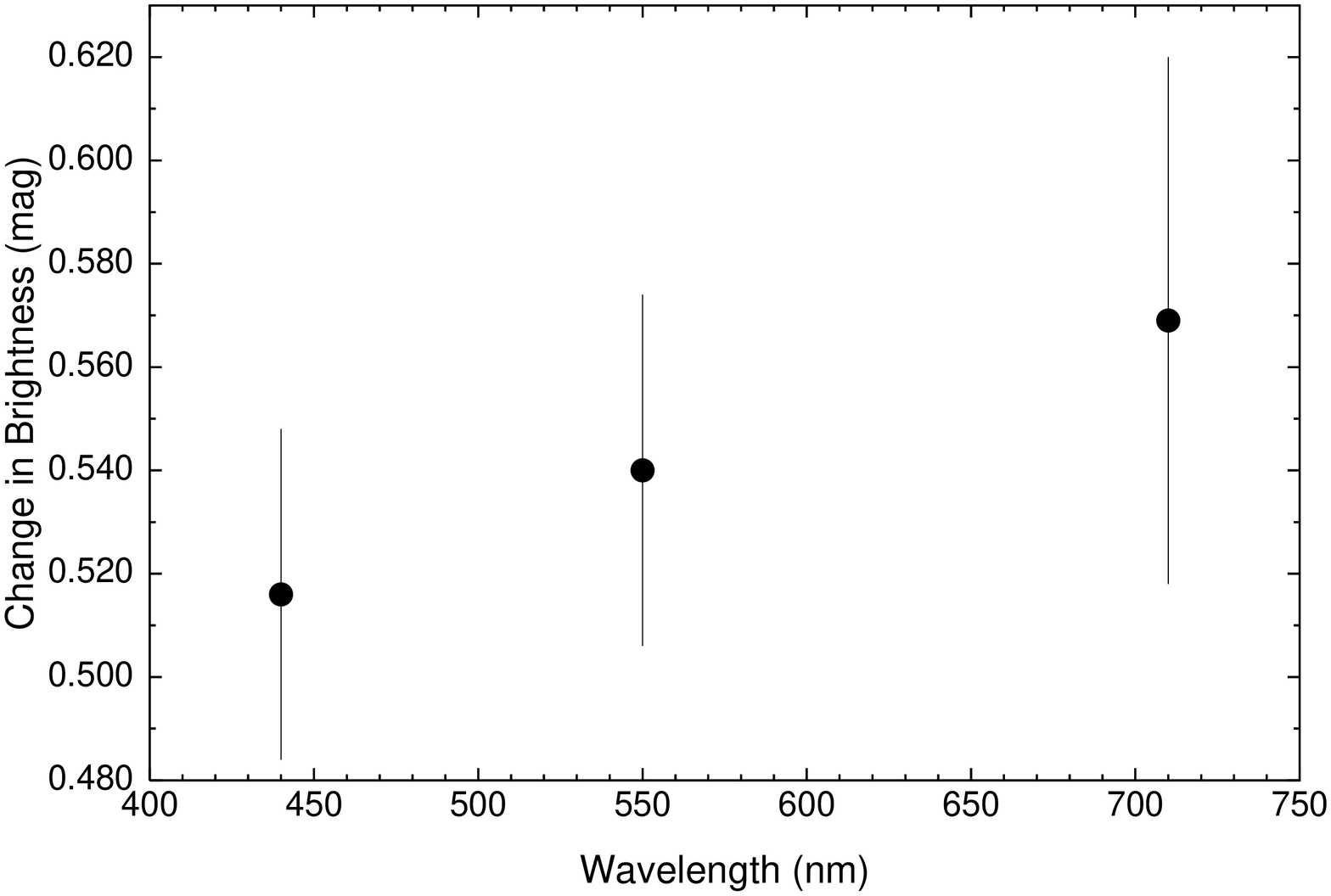}
\figcaption{Optical color behavior of CTA 102 observed in 2005.
These figures display the difference in flare amplitude across the three bands in flux (top) and magnitudes (bottom).}
\end{figure}

\begin{figure}
\includegraphics[angle=270, scale=0.42]{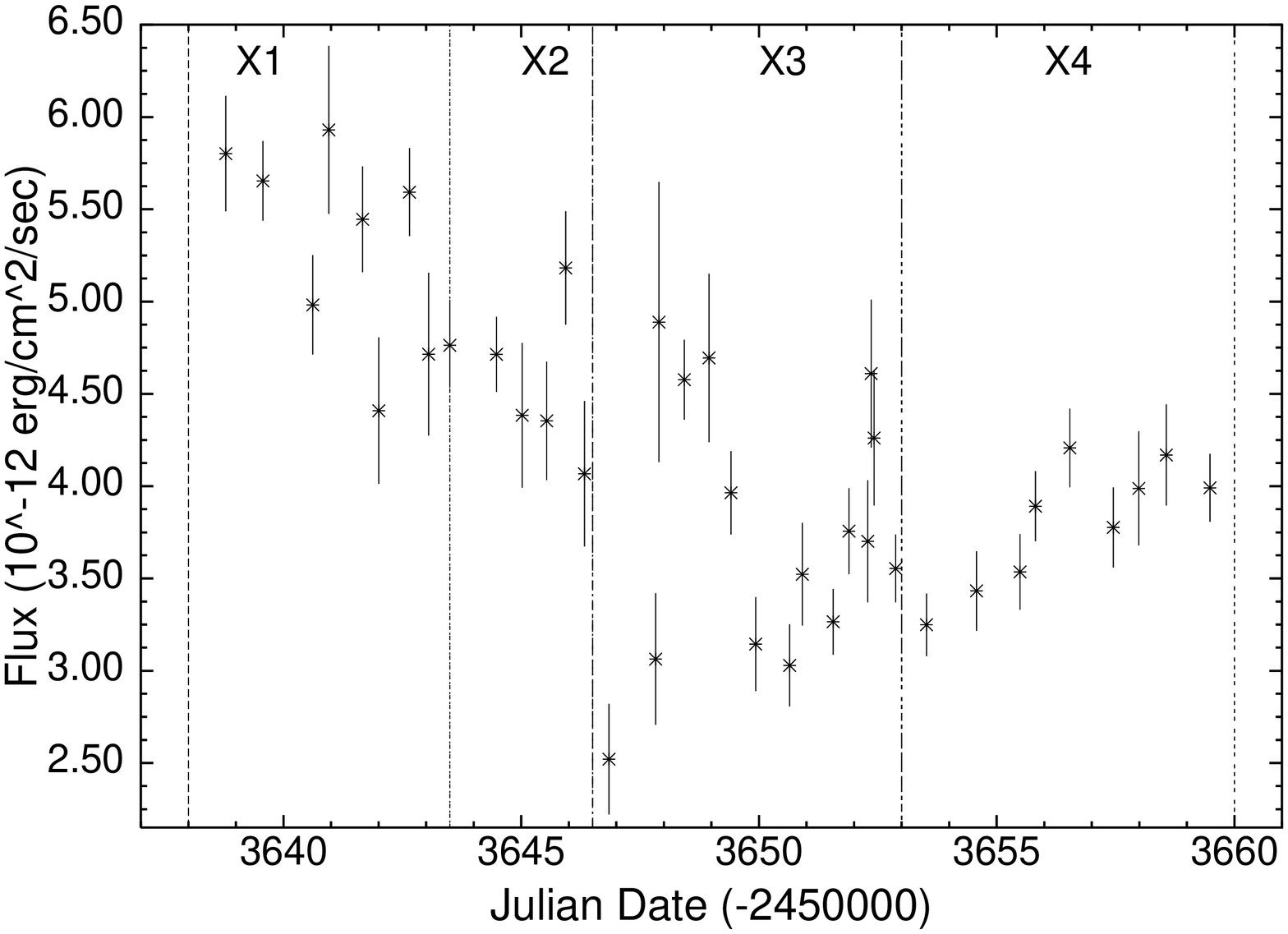}\\
\includegraphics[angle=270, scale=0.40]{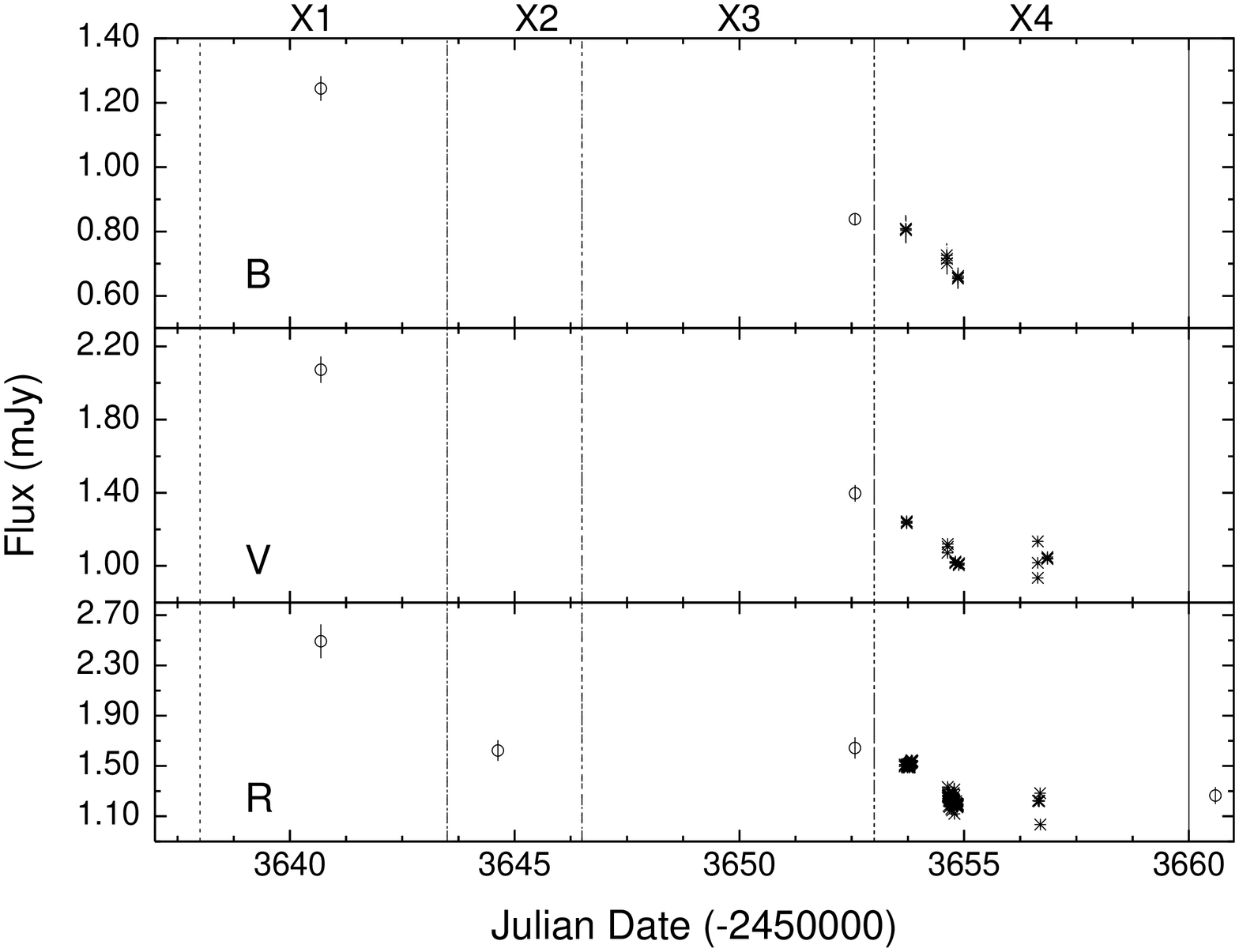}
\figcaption{Simultaneous \emph{RXTE} and R-band observations of CTA 102, with labels X1, X2, X3, and X4 indicating time bins used for generating SEDs. 
See Figure 7 for the key to the optical data symbols. }
\end{figure}

\begin{figure}
\centering
\includegraphics[angle=270, scale=0.40]{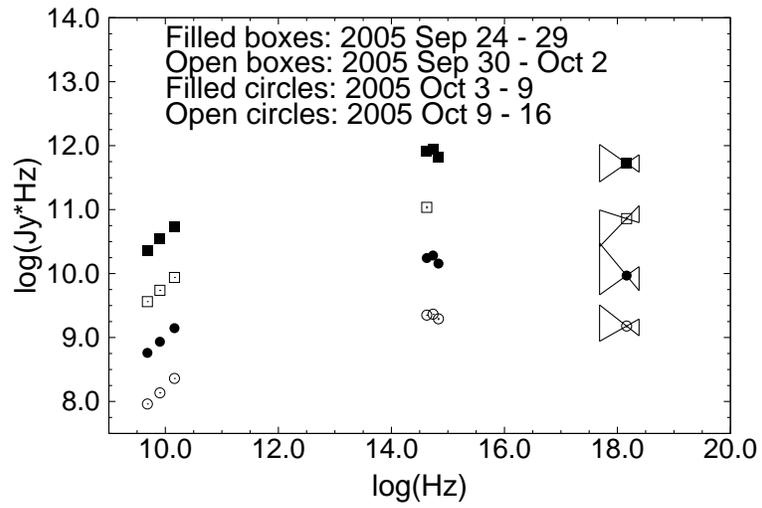}
\caption[seds]{SEDs from temporally correlated 2005 observations of CTA 102. 
The SED from 2005 September 24-29 is plotted at the correct location, with subsequent SEDs vertically offset downward. }\label{seds}
\end{figure}


\begin{thebibliography}{}
\bibitem[Aller et al.(1985)]{all85} Aller, H. D., Aller, M. F., Latimer, G. E., \& Hodge, P. E. 1985, \apjs, 59, 513
\bibitem[Allen et al.(2001)]{cox01} Allen, C. W. et al. 2001, Allen's Astrophysical Quantities (4th ed.; New York: Springer)
\bibitem[Balonek \& Kartaltepe(2002a)]{bal02a} Balonek, T. J. \& Kartaltepe, J. S. 2002a, BAAS, 34, 669
\bibitem[Balonek \& Kartaltepe(2002b)]{bal02b} Balonek, T. J. \& Kartaltepe, J. S. 2002b, BAAS, 34, 1109
\bibitem[Barbieri et al.(1978)]{bar78} Barbieri, C., Romano, G., \& Zambon, M. 1978, \aaps, 31, 401
\bibitem[Beckerman(1997)]{bec97} Beckerman, E. 1997, in Proc. 1997 Undergraduate Symp. on Research in Astronomy, ed. P. J. Benson, T. Balonek, \& J. D'Amore, 14
\bibitem[Benitez \& Ramirez(2006)]{ben06} Benitez, E. \& Ramirez, A. 2006, in ASP Conf. Ser. 350, Blazar Variability Workshop II: Entering the GLAST Era, ed. Miller, H. R., Marshall, K., Webb, J. R., \& Aller, M. F. (San Francisco, CA: ASP), 71
\bibitem[B{\"o}ttcher \& Reimer(2004)]{bot04} B{\"o}ttcher, M. \& Reimer, A. 2004, \apj, 609, 576
\bibitem[Edelson \& Krolik(1988)]{ede88} Edelson, R. A. \& Krolik, J. H. 1988, \apj, 333, 646
\bibitem[Gu et al.(2006)]{gu06} Gu, M. F., Lee, C. U., Yim, H. S., \& Fletcher, A. B. 2006, \aap, 450, 39
\bibitem[Gupta \& Joshi(2005)]{gup05} Gupta, A. C. \& Joshi, U. C. 2005, \aap, 440, 855
\bibitem[Harris \& Roberts(1960)]{har60} Harris, D. E., \& Roberts, J. A. 1960, \pasp, 72, 237
\bibitem[Howard et al.(2004)]{how04} Howard, E. S., Webb, J. R., Pollock, J. T., \& Stencel, R. E. 2004, \aj, 127, 17
\bibitem[Hunstead(1972)]{hun72} Hunstead, R. W. 1972, \aplett, 12, 193
\bibitem[Kataoka et al.(2008)]{kat08} Kataoka, J. et al. 2008, \apj, 672, 787
\bibitem[Mangalam \& Wiita(1993)]{man93} Mangalam, A. V. \& Wiita, P. J. 1993, \apj, 406, 420
\bibitem[Maraschi et al.(1986)]{mar86} Maraschi, L., Ghisellini, G., Tanzi, E. G., \& Treves, A. 1986, \apj, 310, 325
\bibitem[Miller(1981)]{mil81} Miller, H. R. 1981, \apj, 244, 426
\bibitem[Miller et al.(1989)]{mil89} Miller, H. R., Carini, M. T., \& Goodrich, B. D. 1989, Nature, 337, 627
\bibitem[Miller \& Noble(1996)]{mil96} Miller, H. R. \& Noble, J. C. 1996, in ASP Conf. Ser. 110: Blazar Continuum Variability, ed. H. R. Miller, J. R. Webb, \& J. C. Noble (San Francisco, CA: ASP), 17
\bibitem[Moore \& Stockman(1981)]{moo81} Moore, R. L. \& Stockman, H. S. 1981, \apj, 243, 60
\bibitem[Nikolashvili \& Kurtanidze(2005)]{nik05} Nikolashvili, M. G. \& Kurtanidze, O. M. 2005, Mem. Soc. Astron. Ital., 76, 55
\bibitem[Osterman Meyer et al.(2008)]{aom08} Osterman Meyer, A. et al. 2008, \aj, 136, 1398
\bibitem[Pica et al.(1988)]{pic88} Pica, A. J., Smith, A. G., Webb, J. R., Leacock, R. J., Clements, S., \& Gombola, P. P. 1988, \aj, 96, 1215
\bibitem[Racine(1970)]{rac70} Racine, R. 1970, \apjl, 159, L99
\bibitem[Raiteri et al.(1998)]{rai98} Raiteri, C. M. et al. 1998, \aaps, 130, 495
\bibitem[Raiteri et al.(2007)]{rai07} Raiteri, C. M. et al. 2007, \aap, 473, 819
\bibitem[Raiteri et al.(2008)]{rai08} Raiteri, C. M. et al. 2008, \aap, 485, L17
\bibitem[Sandage \& Wyndham(1965)]{san65} Sandage, A. \& Wyndham, J. D. 1965, \apj, 141, 328
\bibitem[Schmidt(1965)]{sch65} Schmidt, M. 1965, \apj, 141, 1295
\bibitem[Soldi et al.(2008)]{sol08} Soldi, S. et al. 2008, \aap, 486, 411
\bibitem[Stalin et al.(2004)]{sta04} Stalin, C. S., Gopal-Krishna, Sagar, R., \& Wiita, P. J. 2004, JA\&A, 25, 1
\bibitem[Tavecchio et al.(2000)]{tav00} Tavecchio, F. et al. 2000, \apj, 543, 535
\bibitem[Tornikoski et al.(1999)]{tor99} Tornikoski, M., Terasranta, H., Balonek, T. J., \& Beckerman, E. 1999, in ASP Conf. Ser. 159, BL Lac Phenomenon, ed. L. O. Takalo \& A. Silanp\"{a}\"{a} (San Francisco, CA: ASP), 307
\bibitem[Ulrich et al.(1997)]{ulr97} Ulrich, M.-H., Maraschi, L., \& Urry, C. M. 1997, \araa, 35, 445
\bibitem[Urry \& Padovani(1995)]{urr95} Urry, C. M., \& Padovani, P. 1995, \pasp, 107, 803
\bibitem[Wagner \& Witzel(1995)]{wag95} Wagner, S. J. \& Witzel, A. 1995, \araa, 33, 163
\bibitem[Welsh(1999)]{wel99} Welsh, W. F. 1999, \pasp, 111, 1347
\bibitem[Wyndham(1965)]{wyn65} Wyndham, J. D. 1965, \aj, 70, 384
\end{thebibliography}
\end{document}